\documentclass[aps,prl,twocolumn,nofootinbib,longbibliography,superscriptaddress,tightenlines]{revtex4-1}

\usepackage{amsmath,amsthm,latexsym,amssymb,amsfonts}
\usepackage{graphicx,color}
\usepackage{hyperref,natbib}

\def\be{\begin{equation}}
\def\ee{\end{equation}}
\def\ba{\begin{eqnarray}}
\def\ea{\end{eqnarray}}

\newcommand\nn{\nonumber}
\newcommand{\q}{\quad}

\begin{document}

\title{Decorated tensor network renormalization\\ for lattice gauge theories and spin foam models}

\author{Bianca Dittrich} 
\affiliation{Perimeter Institute for Theoretical Physics,\\ 31 Caroline Street North, Waterloo, Ontario, Canada N2L 2Y5}
\author{Sebastian Mizera} 
\affiliation{Perimeter Institute for Theoretical Physics,\\ 31 Caroline Street North, Waterloo, Ontario, Canada N2L 2Y5}
\affiliation{Girton College, University of Cambridge,\\ Huntingdon Road, Cambridge CB3 0JG, United Kingdom}
\author{Sebastian Steinhaus}
\affiliation{Perimeter Institute for Theoretical Physics,\\ 31 Caroline Street North, Waterloo, Ontario, Canada N2L 2Y5}

\begin{abstract}
{ Tensor network techniques have proved to be powerful tools that can be employed to explore the large scale dynamics of lattice systems. Nonetheless, the redundancy of degrees of freedom in lattice gauge theories (and related models) poses a challenge for standard tensor network algorithms. We accommodate for such systems by introducing an additional structure decorating the tensor network. 
This allows to explicitly preserve the gauge symmetry of the system under coarse graining and straightforwardly interpret the fixed point tensors.  We propose and test (for models with finite Abelian groups) a coarse graining algorithm for lattice gauge theories based on decorated tensor networks. We also point out that decorated tensor networks are applicable to other models as well, where they provide the advantage to give immediate access to certain expectation values and correlation functions. 
%Using this novel information encoded in the decoration might eventually lead to new methods incorporating both analytical and numerical techniques. 
%We also discuss possible extensions as well as applicability to quantum gravity models.
}
\end{abstract}

\maketitle

\section{Introduction}

Tensor network algorithms \cite{jap1,jap2,levin,gu-wen,beijing} have become a well used method for the coarse graining of classical partition functions.  In particular for models in which Monte Carlo methods are not available (due to sign problems for fermionic systems or due to a priori complex amplitudes as in spin foams\footnote{Spin foams are path integrals for quantum gravity, defined without a Wick rotation. They can be understood as generalized gauge theories, see for instance \cite{foams1,foams2,foams3,finitesf,holonomy1}.}) tensor network algorithms may be the only way to explore the phase structure of a given system. We are in particular interested in the phase structure of spin foams, candidate models for quantum gravity. For these models the key open question is the recovery of a dynamics describing continuum manifolds in the limit involving many building blocks. To make progress on this question it is essential to understand the phase structure of spin foam models. 

Spin foams are generalized lattice gauge theories. In this work we will present the first tensor network (renormalization) algorithm, applicable to lattice gauge theories (including spin foams) that avoids the summation over unphysical gauge degrees of freedom.  This makes the algorithm much more efficient than the one proposed recently in \cite{meurice}. We will test this algorithm  explicitly for the 3D Ising gauge model.  Additionally the tensor network algorithm we propose, generalizes the notion of tensor networks to a new type of structure that we call `decorated tensor networks'. This new structure has the advantage to allow for a convenient access to expectation values of certain observables (encoded in the decorations). 

Let us comment on the current state of the art in tensor network renormalization in particular with regard to lattice gauge theories. Tensor network algorithms \cite{jap1,jap2,levin,gu-wen,beijing} have been successfully used to evaluate classical partition functions in particular in two dimensions, less so in three dimensions. Very recent work \cite{vidal-evenbly} designed a tensor network algorithm incorporating `entanglement filtering' which allows for a very efficient exploration of (2D) models also at phase transitions.  

As mentioned an advantage of tensor network algorithm is that they can be applied to systems where sign problems or complex amplitudes prevent the use of Monte Carlo methods. This applies to (for instance lattice gauge) systems with fermions as well as spin foams. In the latter case no other systematic methods for coarse graining are known.

However not much work has been done concerning the development of tensor network renormalization algorithms for lattice gauge theories in 3D (and 4D). A tensor network representation for lattice gauge theories and spin foam models appeared in \cite{eckert1,eckert2} and later also in \cite{meurice}. The recent work \cite{meurice} proposed, based on this tensor network representation, an algorithm (which has not been tested). As we will explain later, this representation is however not  economic in terms  of the `size' of the initial tensor, that is the range of its indices. In the literature this is denoted as the bond dimension of the tensor. As the computational resources required grow exponentially with this bond dimension, it is vital to avoid redundancies. The method presented in this work leads to a much more efficient algorithm.  

More activity has been spent on an alternative usage of tensor networks in lattice gauge theories, namely to provide a variational ansatz for low energy wave functions  \cite{vidalgauge, tagz, osborn, ciracgauge}. Here the issue is again to find an efficient representation (`efficient' for the purpose of computing the expectation value of certain observables) of these wave functions and to avoid an over-parametrization due to the gauge symmetry. These works consider $(1+1)$ or $(2+1)$ dimensional systems. 

In contrast to lattice gauge theories the phase structure of spin foams is largely unknown and tensor network techniques are so far the only approach to systematically coarse grain such models. We will shortly review the progress towards reaching a phase diagram for spin foams and the particular relevance of developing tensor network based methods for these models in the appendix.

All (numerical) tensor network algorithms have in common to rely eventually on the summation over the variables associated to sites or edges etc. of the lattice. In tensor network terms such models have a finite bond dimension. There is thus an important limitation, which is that the summation range needs to be (a priori) finite. This is the case for lattice gauge theories with a finite group or with a quantum group at root of unity. The latter case is particularly relevant for spin foams as models with a quantum group are proposed to incorporate the cosmological constant \cite{qgroupmodels,qgroupmodels1,qgroupmodels2,qgroupmodels3,qgroupmodels4,qgroupmodels5,qgroupmodels6,qgroupmodels7}, and thus tensor network methods are indeed applicable to such spin foam models.

Nevertheless there are different proposals how to deal with Lie groups such as $U(1)$ or $SU(2)$, that would lead to infinite bond dimension (in the standard version of lattice gauge theories). One is to use Monte Carlo methods for the summation over variables (encoded in the contraction of indices for the tensor network) \cite{vidalMC}, another is  to incorporate a cut-off that would capture the relevant degrees of freedom \cite{phi4,tagz}. The latter approach would however limit the regimes for which reliable results can be expected.\footnote{In, say $SU(2)$ lattice gauge theory at strong coupling the weights peak for small (spin) representation labels $j$. Thus a cut-off might be applicable. However in the weakly coupled regime the weights are rather flatly distributed over the representation labels. Increasing the cut—off one can test smaller and smaller couplings and eventually match the results (e.g. expectation values of observables) to the perturbative expansion. A similar strategy is employed in Monte-Carlo simulations where testing the small coupling regime at small lattice length (or large number of lattice sites) is also challenging.} Replacing the Lie groups with a quantum group (at root of unity) leads to a naturally induced cut--off on the initial bond dimension required. 
See also \cite{bern,bern1} for a definition of lattice gauge models  (gauge magnets) with Lie groups as gauge groups but with finite bond dimensions.  

This discussion shows that designing efficient algorithm (which is expressed as the scaling of the computational time in terms of the bond dimension) is rather important. As mentioned quantum group based models lead to a deformed symmetry group with a natural cut--off on the initial bond dimension. For spin foams such models describe gravity with a cosmological constant. For small cosmological constant one however needs a large initial bond dimension. 
The decorated tensor network models introduced here might provide an additional alternative to deal with infinite bond dimension as they might eventually allow for a mixture of analytical and numerical techniques.

{\bf Outline:} 
The paper is organized as follows:  Although we have applications to lattice gauge theories and spin foam models with non--Abelian structure groups in mind, we will concentrate in the main text on Abelian groups. This allows us to emphasize the main features of the new decorated tensor network algorithm more clearly. We discuss in the appendix  the structure of theories with non--Abelian groups and give a short outline of the coarse graining algorithm for non--Abelian groups in the main text. The details of this algorithm, which includes the new feature of protecting a non--Abelian gauge  symmetry, will appear elsewhere. (See  for instance \cite{vidalsymm,merce,qgroup} for a discussion of symmetry protecting algorithms for global symmetries.)

We therefore start with a review of the structure of Abelian lattice gauge theories and discuss different tensor network representations. (We defer the discussion for the non--Abelian case to the appendix.) This discussion motivates the introduction  of a new type of tensor networks, called decorated tensor networks. These preserve explicitly the Gau\ss~constraints representing the gauge symmetry under coarse graining. We explain a 3D coarse graining algorithm for decorated tensor networks first in the lowest order approximation and then different versions of higher order approximations. 
We also  test these versions for the 3D Ising  gauge theory, i.e.\ lattice gauge theory with ${\mathbb Z}_2$ gauge group.\footnote{This theory is dual to the (non--gauge) Ising model \cite{savit}, which one can see by solving the Gauss constraints. However such a duality transform is not available for (generalized) non--Abelian gauge theories, to which we hope to apply decorated tensor networks eventually.  We therefore keep the Gauss constraints explicit.} We also briefly comment on extensions of this algorithm to non--Abelian gauge theories. Afterwards we point out that decorated tensor networks are also applicable to other models beside lattice gauge theories. We present a decorated tensor network  algorithm applicable to Ising--like systems which preserve (part of) the original spin variables of the system, such that correlation functions can be straightforwardly calculated. We close with a discussion of the algorithms and results.

\section{Representations for (Abelian) lattice gauge theories}\label{lgt}

Here we will shortly review representations of lattice gauge theories and the encodings of the corresponding partition functions into tensor networks.

Lattice gauge models feature a partition function
\ba\label{1}
Z&=& \frac{1}{|G|^\sharp e}\sum_{g_e} \, \prod_f \omega \left( \vec{\prod}_{e \subset f} g_e \right) \, ,
\ea
where $g_e$ denotes group elements attached to the (oriented) edges $e$ of a lattice, $f$ denotes an (oriented) face or plaquette of the lattice and $h_f= \vec{\prod}_{e \subset f} g_e$  the face holonomy, that is the oriented product of group elements associated to the edges belonging to this face.  $|G|$ denotes the cardinality of the group, and $\sharp e$ the number of edges in the lattice. The face weights $\omega$ have to be class functions, i.e. $\omega(ghg^{-1})=\omega(h)$, for the partition function to be invariant under gauge transformations $g_e\rightarrow h^{-1}_{s(e)} g_e h_{t(e)}$. Here $h_v$ denote  group elements, acting as gauge parameter, associated to the vertices of the lattice, $s(e)$ and $t(e)$ the source and target vertex of the edge $e$ respectively.

We assumed a finite structure group in (\ref{1}), for application to Lie groups we have to replace the sum in (\ref{1}) by an integral.  Later we will perform a duality transformation, in which the partition function appears as a finite/infinite sum for finite/ compact Lie groups. Moreover in this representation we can also easily deal with the quantum group case.

The gauge transformations act at the vertices by transforming the group elements $g_e$ of the adjacent edges: $g_e\rightarrow h^{-1}_{s(e)} g_e h_{t(e)}$ where $h_v$ denotes the (group--valued) gauge parameter associated to the vertex $v$, with $v$ being the source $s(e)$ and target vertex $t(e)$ of $e$ respectively. Gauge transformations leave the weights of the partition function invariant, the weights are thus constant along orbits of the gauge action. The summation in (\ref{1}) includes therefore redundant parts, namely the summation over configurations in one and the same gauge orbit. As tensor network methods aim to explicitly evaluate the partition function through a summation, it is important to avoid these redundant parts in a tensor network representation in order to obtain a more efficient algorithm.

Let us now specify to Abelian structure groups $\mathbb{Z}_K$ (so that $g \in \{0,\ldots, K-1\}$, with addition modulo $K$ as group multiplication). We give the details for lattice gauge theories with non--Abelian structure groups and spin foams in the appendix.  

Although  the representation (\ref{1}) features variables based on edges, as in tensor networks, the weights are actually associated to the faces and not to the vertices. Additionally the gauge symmetry leads to an over-parametrization of the system, which should be avoided.  To achieve a more appropriate representation we perform a duality transformation (or strong coupling / high temperature expansion), which rewrites the partition function using a (group) Fourier transform for the weights (see for instance \cite{savit}): 
\ba\label{2}
\omega(g)=\sum_k \tilde \omega(k) \chi_k(g) \, , 
\ea
 where $k \in  \{0,\ldots, K-1\} = \mathbb{Z}_K$ labels the irreducible unitary representations of the group $\mathbb{Z}_K$ and
 $\chi_k(g)= \exp(\frac{2\pi i}{K} g\cdot k)$   denoting the character of the representation.

Using (\ref{2}) in (\ref{1}) for every face weight and the multiplicative property of the characters $\chi_K(g_1 g_2)=\chi_K(g_1) \cdot \chi_K(g_2)$ we can rewrite the partition function into
\ba
Z&=&  \sum_{k_f}  \prod_f  \tilde \omega(k_f)  \,\, \prod_e \left(  \frac{1}{K}\sum_{g_e}  \prod_{ f \subset e}   \chi_{ k_f}(g_e^{o(e,f)} )\right) 
\ea 
where $o(e,f)=-1$ if face and edge orientation disagree and $o(e,f)=+1$ if these agree. One now integrates out the group elements, which gives a Kronecker--delta for each edge:
\ba\label{AbZ}
Z&=&\sum_{k_f}  \prod_f \tilde \omega(k_f)  \,   \prod_e  \delta\left(\sum_{f \supset e} o(e,f) \, k_f\right)  .
\ea
These delta functions impose the so--called Gauss constraints, stating that the representations around the faces meeting in one edge sum up to zero.

\begin{figure}
\begin{center}
\includegraphics[width=0.39\textwidth]{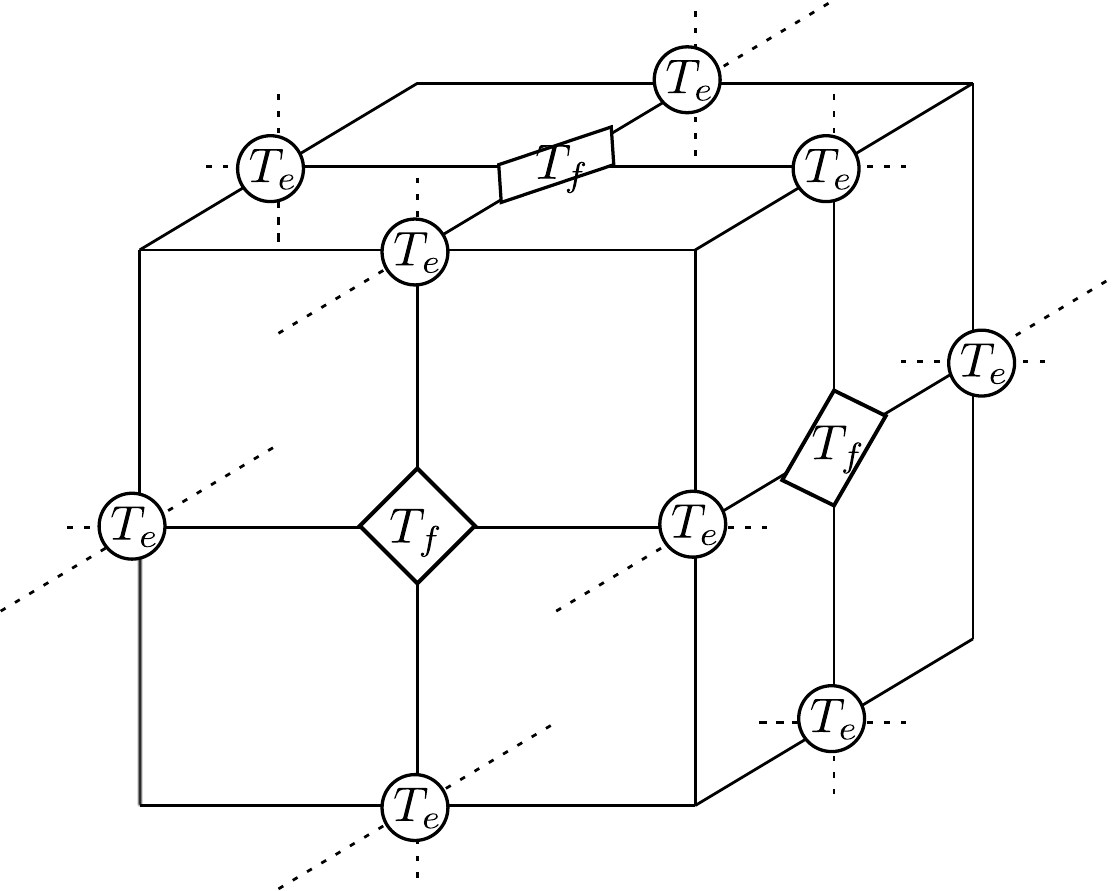}
\caption{Tensor network representation of a spin foam defined on a cubical lattice. The tensors $T_e$ on the edges capture the intertwiner degrees of freedom, while the tensors $T_f$ on the faces are an auxiliary structure. For a given face $f$ they ensure that all edge tensors $T_e$ with $e \subset f$ carry the same representation label $\rho_f$ assigned to that face.
\label{fig:sf-tensor}}
\end{center}
\end{figure}

The Gauss constraints can be understood to follow from the gauge symmetry of the action. Instead of having redundant variables we now have constrained variables. An effective tensor network algorithm would rather solve the Gauss constraints instead of leaving the delta-function weights in the partition function and summing over all variables. 

Indeed the full duality transformation (for instance reviewed in \cite{savit}) does solve the Gauss constraints by introducing  potential (variables) for the variables $k_f$.  For a 3D lattice these potential variables  $l_c$ are located in the centre of the cubes, so that $k_f=l_{c_{f+}}-l_{c_{f-}}$  where $c_{f+}$ is the adjacent cube to $f$ with positive induced orientation on $f$ and ${c_{f-}}$ is the one with negative induced orientation. Substituting  the variables $k_f$ with $k_f(l_c)$, the Gauss constraints are now automatically satisfied. The partition function is now a (generalized Ising) $\mathbb{Z}_K$ model with $\mathbb{Z}_K$ as global symmetry group and dual weights $\tilde \omega$. 

However this strategy is only available for the 3D Abelian models and does not apply to non-Abelian ones or to spin foams. (In 4D the duality transform maps an Abelian gauge theory to an Abelian gauge theory.)  As we are interested in methods that are also applicable to non-Abelian structure groups, we will not pursue this option here.

Let us now review possible tensor network representations of the partition function (\ref{AbZ}). Although we have now variables $k_f$ associated to faces, this form of the partition function (\ref{AbZ}) can be brought into the form of a tensor network. To this end one understands the Kronecker-Delta associated to a given edge  
$\delta(\sum_{f \supset e} o(e,f) \, k_f ) ={\mathbf P}_e( \{k_f\})$ as tensor\footnote{These tensors are much more non--trivial for non-Abelian gauge theories.} with indices $k_f$ (for faces adjacent to the edge $e$). One associates these tensors to the midpoints of each edge and introduces auxiliary tensors (again given by Kronecker--Deltas) to the midpoints of the faces, that make sure, that each tensor $({\mathbf P}_e )$ around a given face $f$, sees the same representation label $k_f$, see also figure \ref{fig:sf-tensor}. 
The face weights $\tilde \omega_f$ can be  either multiplied to these auxiliary tensors or distributed to the adjacent edges, that is each tensor ${\mathbf P}_e( \{k_f\})$ is multiplied with a factor $(\tilde \omega_f)^{1/4}$ for each of the adjacent faces $f$. 
This representation appeared in \cite{eckert1,eckert2} and was also used for the coarse graining algorithm presented in  \cite{meurice}.

However, this representation (\ref{AbZ}) is rather inefficient if it comes to coarse graining algorithms. The reason is that the Gauss constraints are put as Kronecker--Delta's into the tensors instead of solving them.  One thus sums over much more variables than is actually necessary. 

We will introduce therefore yet another representation in which the weights will be placed on the vertices of the lattice and is then called vertex amplitude $A_v$.  This procedure  generalizes  to non-Abelian gauge theories and spin foams. For these theories it involves a splitting of the tensors $({\mathbf P}_e )$ to tensors associated to half edges and a contraction of all half edge tensors meeting at a vertex. For the Abelian case the procedure simplifies drastically: we just have to double the Kronecker--Delta's associated to each edge so that we have again one Kronecker--Delta per half edge. We then summarize all the Kronecker-Delta's associated to the half edges meeting at a vertex to one vertex amplitude.  We also distribute the weight factors  to the vertices, by associating $\tilde \omega_f^{1/4}$ to a vertex for each face adjacent to that vertex. The vertex amplitudes are then given as
\ba\label{AbAv}
A_v = \prod_{f \supset v} \tilde \omega_f^{1/4}  \prod_{e \supset v}   \delta\left(\sum_{f \supset e} o(e,f) \, k_f\right)  \q .
\ea

As we will see, an advantage of this representation is that the Gauss constraints can be solved locally at each vertex. As in a tensor network we do now have amplitudes associated to vertices, the summation is however over variables $k_f$ associated to faces, where each variable appears in four vertex amplitudes (whereas in a tensor network each variable appears in only two tensors). 

\begin{figure}
\begin{center}
\includegraphics[width= 0.48\textwidth]{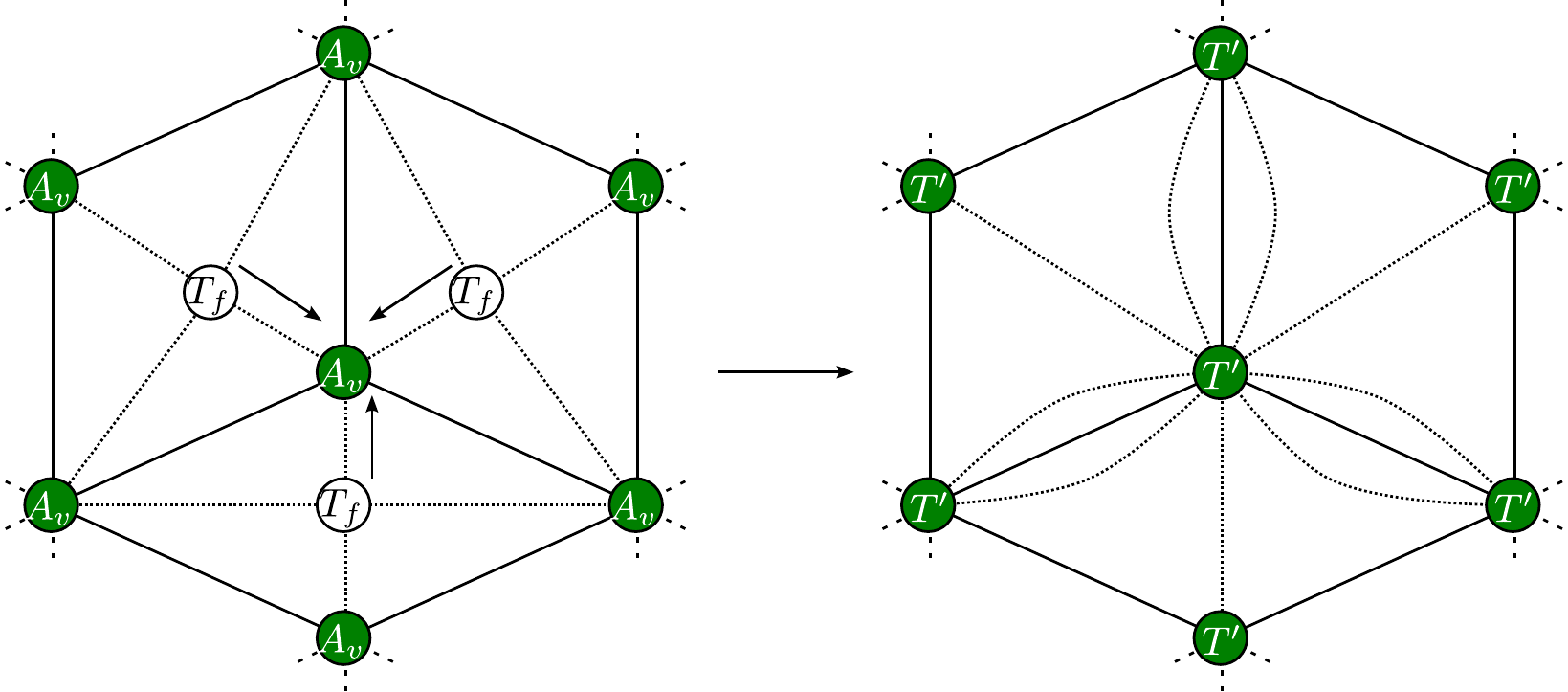}
\caption{Definition of a homogeneous, yet anisotropic tensor network from the vertex amplitude representation. By absorbing the auxiliary tensors in the positive quadrants into the vertex, one obtains a cubical tensor network, with tensors $T'$, with additional diagonal edges (some of the auxiliary edges can be combined with the original cubical edges, increasing the bond dimension). The regular cubical edges get equipped with additional data, namely the representation on the face.
\label{fig:new-rep}}
\end{center}
\end{figure}

This could again be dealt with by introducing auxiliary tensors put into the middle of the faces. One can then attempt to group tensors of this inhomogeneous network into a homogeneous one, resulting in tensors with more edges and also higher initial bond dimension (see  \cite{meurice} where this has been applied to the representation (\ref{AbZ})). For a cubical 3D lattice a possible grouping is depicted in figure \ref{fig:new-rep} and would result in  12--valent tensors, which are also connected via diagonal edges.

Alternatively\footnote{We thank J. Hnybida for pointing this out.}, one can actually copy a given face index $k_f$ to all edges adjacent to this face. To enforce equality between the copied indices we have to multiply the vertex amplitudes  with Kronecker deltas.  Each edge would a priori carry four variables $k_{f \supset e}$, however the Gauss constraints would reduce this to three. This would then still give for a structure group ${\mathbb Z}_K$ and a 3D hyper cubical lattice a bond dimension $K^3$. This again is a rather inefficient way to proceed, which results from the need to introduce additional Kronecker--Delta's at the vertices.

\section{Decorated tensor network renormalization}

We have discussed  in the last section different possibilities to rewrite a lattice gauge theory partition function into tensor network form. However the problem is that summations do not appear naturally in the form of tensor contractions (i.e. as a sum over a variable which appears only in two weights). Thus one either copies variables or introduces auxiliary tensors in the form of Kronecker--delta's, which renders the tensor network representation inefficient.

We  therefore propose to generalize the concept of tensor networks to a structure that we call `decorated tensor networks'.  These decorated tensors allow to keep the original type of variables and structure of summation at each coarse graining step. (Note that the variables are `blocked' in each coarse graining step.)  In our example these variables are given by the representation labels $k_f$ associated to (larger and larger) faces. 
At lowest possible truncation, which gives the minimal possible bond dimension, one indeed deals only with these variables and the original structure of summation. One can increase the truncation by introducing a proper tensor network structure. This tensor network would carry the fine grained variables, needed for the higher order approximation, which are not captured by the blocked variables. On the other hand the original variables (here $k_f$) appear as `decoration' of this tensor network. Summing over all tensor network indices one is left with an effective partition function in the (blocked) variables $k_f$, with in general non-local (i.e. non nearest neighbour) couplings induced by this summation. 

Apart from allowing for a more efficient representation of partition functions, in particular at low order in the approximation\footnote{This is quite important as for larger structure groups the lowest possible truncation leads to a larger minimal bond dimension.}, the decorated tensor network algorithm allows an easy access to the expectation values of the (blocked) variables $k_f$. This is different from standard tensor network methods where the determination of expectation values comes with higher costs in the bond dimension.

For the following discussion we found it simpler to transform the representation to the dual lattice.\footnote{For spin foams this dual lattice would give the geometric building blocks and the representation labels give the length (in 3D) of the edges in these building blocks. The 'blocked' variables do then correspond to the length of edges of the glued building blocks.} Vertices  in the original lattice are replaced by cubes in the dual lattice and vice versa. Faces (or plaquettes) in the original lattice are replaced by edges (which cross these faces orthogonally) connecting the new vertices at the centres of the old cubes. 

 Thus `tensors based on vertices' are now replaced by `amplitudes associated to building blocks'. To some degree this is just a change of perspective, see also \cite{dittcyl}, but it allows for more flexibility in designing (and imagining) algorithms. For a cubical 3D lattice this means that we now associate amplitudes to cubes. Variables which have been associated to faces, are now identified with the edges $k_f \rightarrow k_e$ .%, and the intertwiner labels are associated to the faces $\iota_e \rightarrow \iota_f$, see figure \ref{fig:decorated}. 
 %The coupling condition is now also based on the faces, i.e. the representations associated to the edges of a face have to couple to the intertwiner label associated to this face. The coarse-graining / blocking algorithms now proceed by gluing cubes to larger cubes, by summing over the labels associated to bulk edges and bulk faces.

\begin{figure}
\begin{center}
\includegraphics[width=0.35\textwidth]{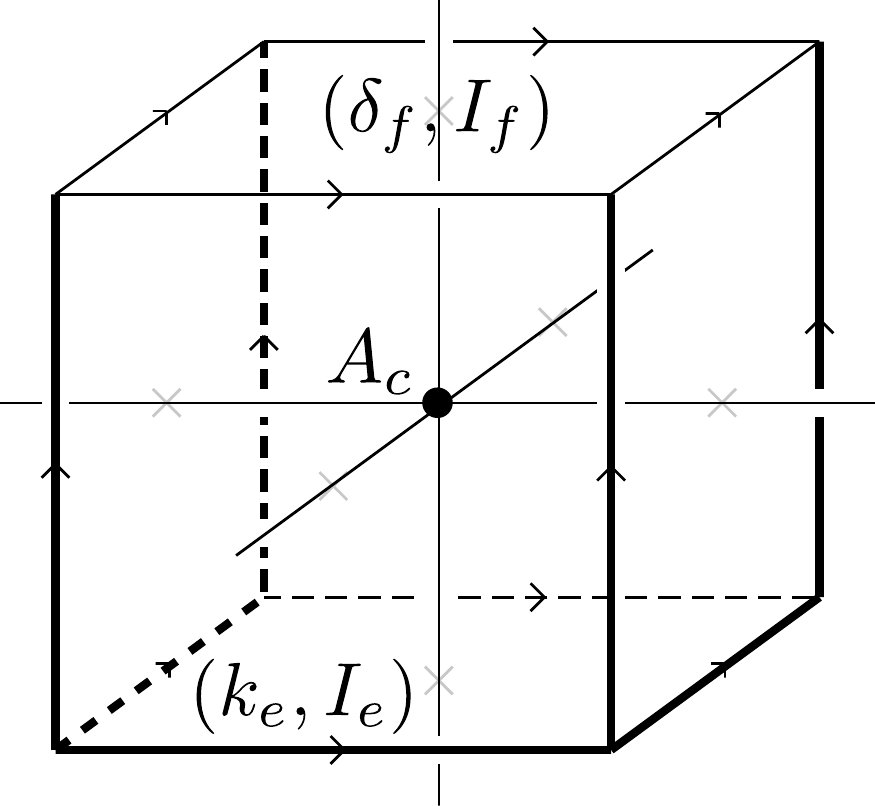}
\caption{
Example of a basic building block of the decorated tensor network. The amplitude tensor $A_c$ is associated to the cube, and contains the information about edge variables, $k_e$ and $I_e$ (appearing for higher order approximation) and face variables $\delta_f$ (these might arise in Abelian models for subdivides faces) and $I_f$ (appearing for higher order approximation).  Bold edges represent a choice of  maximal tree that represents the free variables for Abelian groups. We depict face variables as legs of the central tensor piercing through respective faces. One can think of it as one element of a tensor network decorated with additional edge variables. % Later--on we will introduce additional indices associated to edges $I_e$ and/or faces $I_f$.
\label{fig:decorated}}
\end{center}
\end{figure}

%\subsection{Abelian gauge models in 3D}

%Let us specialize to Abelian gauge models in 3D. These models are dual to Abelian `edge' models with global symmetry \cite{kramers1,kramers2,savit}, as we will see shortly. However, as we want to test algorithms also applicable to non-Abelian gauge models, where such a duality is not available, we will keep the lattice gauge (spin foam) representation of the model.

%For Abelian models, for instance  with structure group ${\mathbb Z}_K$, we denote the representations associated to the (oriented) edges by $k_e$ with $k_e =0, \ldots, K-1$. 

The amplitude associated to a cube is now given as 
\ba\label{initialcubeA}
A_c =\prod_{e \in c} \left( \tilde\omega(k_e) \right)^{1/4} \,  \prod_{f \in c} \delta\left( \sum_{e \subset f} o(e,f) \, k_e    \right) \q .
\ea
The second factor in (\ref{initialcubeA}) describes the Gauss constraints.  These constraints can be solved for each cube in the following way: One fixes an independent set of 7 variables $k_e$  by choosing a maximal tree (a connected subgraph without loops) among the edges of a cube. By solving the Gauss constraints for the 5 edges which are not part of the tree, the cube amplitude does depend only on 7 variables instead of the original 12 variables $k_e$. One can now also omit the Kronecker--delta's in (\ref{initialcubeA}).
If we need to access the variables associated to edges of the tree (called branches), for instance for gluing two cubes by summing over shared variables, one can easily reconstruct these by using the Gauss constraints. 

In the lowest order approximation we will work with these cube amplitudes depending on the seven variables $k_e$. (The coarse graining algorithm will actually introduce an additional subdivision on two opposite faces, so that one actually deals with nine variables.) For higher order approximation we will introduce a tensor network which sits on the lattice dual to the current one (i.e. on the original lattice). That is the tensor network edges will pierce the faces of the cubes -- and thus we associate further variables (tensor network indices) to these faces. The cube amplitude is subsumed into the tensor associated to the cube or alternatively to the vertex in the centre of the cube. The tensor carries the tensor indices (which will be summed over by contraction) as well as the variables $k_e$ which one can understand as additional parameters for this tensor.

\subsection{Algorithm in the leading order approximation}

Let us first convey the spirit of the coarse graining algorithm for the leading order approximation. In this case the algorithm will proceed without the introduction of (tensor network) indices carrying variables in addition to the representation labels $\{k_e\}$.  Thus throughout the algorithm we will keep  the same variable structure (with a small extension of the number of independent variables from 7 to 9) for the cube amplitudes $A_c$. This amplitude will however flow and loose its initial factorizing form. 

The extension in the number of variables for the cube is as follows: Two opposite faces of the cube will be subdivided by four edges into four triangles. Each of the new edges carries an additional variable $k_e$. However we also have four new faces, each carrying an additional Gauss constraint. The four Gauss constraints for a subdivided face imply the Gauss constraint for the very same face, there are therefore only three additional independent constraints. In summary we just add two new independent variables for the two subdivided faces.

\begin{figure}
\begin{center}
\includegraphics[width=0.425\textwidth]{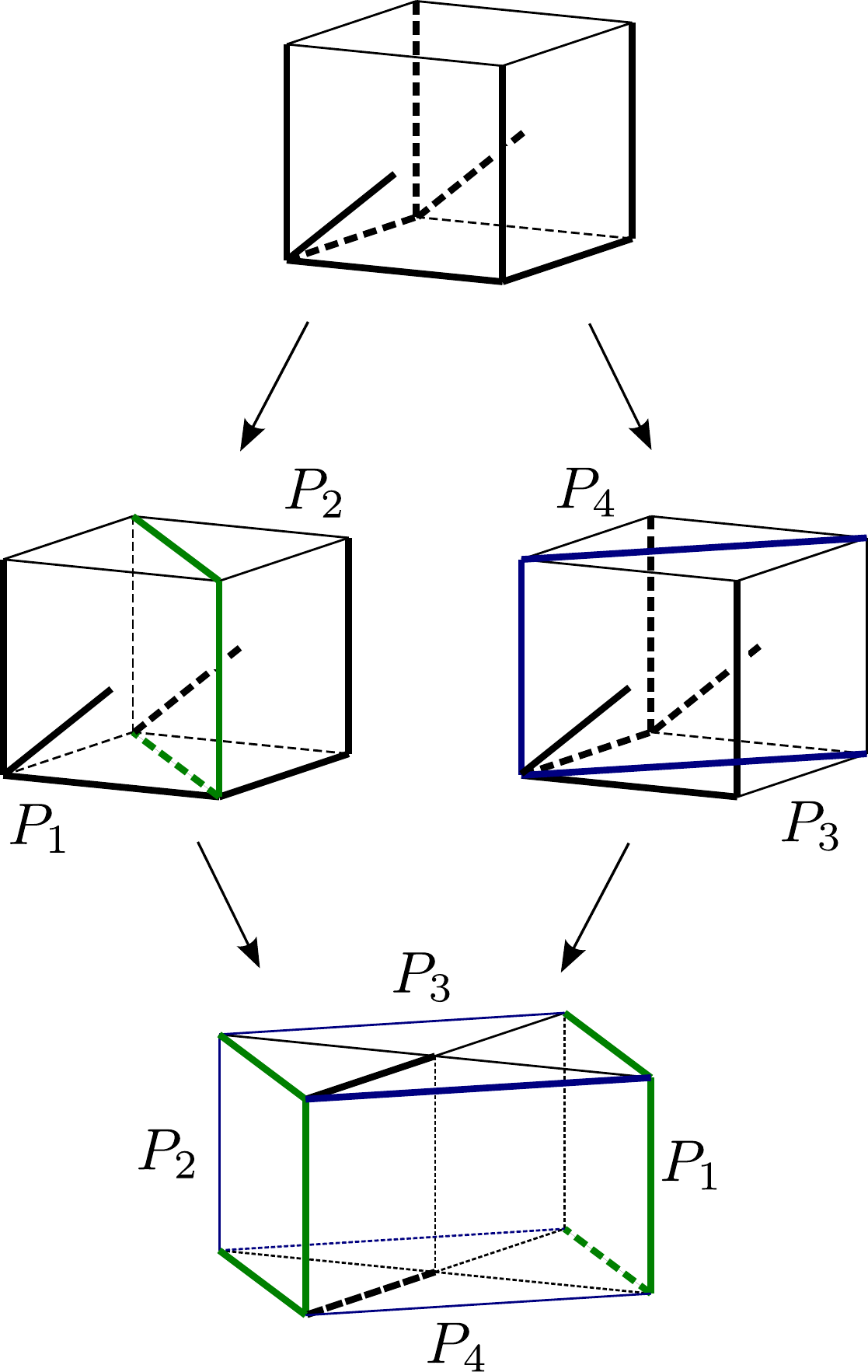}
\caption{Full iteration of the algorithm. The initial cube is split in two ways into prisms, which are then glued back together into a rhomboid again. The following iteration is performed after a rotation into a different plane. In order to complete the iteration cycle we keep an auxiliary half-edge on two faces, which is summed over in the last step. The bold edges indicate the choice of a maximal tree, the coloured edges highlight the edges shared by the split prisms. %For details refer to the text.
\label{fig:3d-splitting}}
\end{center}
\end{figure}

The basic algorithm is a (decorated) 3D generalization of the Gu--Wen algorithm \cite{gu-wen} as shown in  figure \ref{fig:3d-splitting}: We divide each cube into two prisms using a singular value decomposition (SVD). Then four of these prisms are glued to a new cube.
(The new building blocks are actually parallelepipeds or rhomboids, but we will refer to them as cubes.)
We changed the lattice length, say in the plane $(xy)$, by a factor of $\sqrt{2}$, see figure \ref{fig:3d-splitting}. We then rotate the lattice, so that in the next iteration we coarse grain in a plane orthogonal to the previous one.

Let us describe the steps of the algorithm in more detail:\\

{\bf Splitting the cubes into prisms:} To divide each cube into two prisms, we have to introduce diagonals for two faces (opposite to each other) of the cube. Due to the Gauss constraints for the new triangular faces the edges of the subdivided faces determine the representation label for the new edges uniquely. As we introduce two new variables and two additional (independent) constraints, the counting of independent variables does not change.
 
 `Splitting the cube into two prisms' then means that we approximate the amplitude for the cube by a sum over variables appearing in the amplitudes associated for the two prisms $P_A$ and $P_B$:
 \ba\label{splitcube}
&& A_c(\{k_e\}_{e \in {\mathbb A}}, \{k_e\}_{e \in {\mathbb B}}, \{k_e\}_{e \in {\mathbb C}}) \approx  \nn\\
&&\sum_I  A_{P_A} (\{k_e\}_{e \in {\mathbb A}},  \{k_e\}_{e \in {\mathbb C}}, I) \,\, A_{P_B} (\{k_e\}_{e \in {\mathbb B}},  \{k_e\}_{e \in {\mathbb C}}, I) \, .\nn\\
 \ea
 Here $\{k_e\}_{e \in {\mathbb A}}$ denote the independent variables associated to the prism $P_A$, but do not appear in $P_B$. Similarly $\{k_e\}_{e \in {\mathbb B}}$ denote the independent variables associated to the prism $P_B$, but do not appear in $P_A$. The variables $\{k_e\}_{e \in {\mathbb C}}$ appear both in $P_A$ and in $P_B$.  For instance in the left--central panel in figure \ref{fig:3d-splitting} the bold edges carry the independent variables. The three bold edges to the left of the dividing plane carry the $\{k_e\}_{e \in {\mathbb A}}$, the three bold edges to the right of this plane carry the $\{k_e\}_{e \in {\mathbb B}}$ and the three green bold edges in the plane carry the variables $\{k_e\}_{e \in {\mathbb C}}$.
 
The allowed range of the summation index  $I$ on the right hand side of (\ref{splitcube}) determines the order of the approximation of this algorithm. In the lowest possible order one chooses the range to be equal to one, so that  in (\ref{splitcube}) the cube amplitude is approximated by the product of the two prism amplitudes. %Note that the number of singular values  (in the singular value decomposition discussed below) taken into account in each splitting is $K^3$  -- one per index combination  $\{k_e\}_{e \in {\mathbb C}}$.
 
To find the best possible approximation in (\ref{splitcube}) we apply a singular value decomposition (SVD) -- which is a generalization of the better known eigenvalue decomposition. The SVD allows to write a matrix $M_{AB}$ as
\ba
M_{AB} \,=\, \sum_I  U_{AI} \lambda_I V^\dagger_{BI}
\ea
with unitary matrices $U_{AI}$ and $V_{IB}$ and positive singular values $\lambda_I$.

 Note first that this method is only applicable to matrices. Therefore we summarize the variables $\{k_e\}_{e \in {\mathbb A}}$ into an index $A$, the variables $\{k_e\}_{e \in {\mathbb B}}$ into an index $B$ and $\{k_e\}_{e \in {\mathbb C}}$ into $C$. We introduce the matrices  $M^C_{AB}=A_c(A,B,C)$ labelled by the index $C$. 
 
Via the SVD we split the matrices $M^C_{AB}$
 \ba
 M^C_{AB}& \,=\,& \sum_I \left( U^C_{AI} \sqrt{\lambda^C_I} \right) \left( \sqrt{\lambda^C_I} V^\dagger_{BI} \right) \nn\\&\,=:\,& \sum_I (S^{1})^C_{AI} (S^{2})^C_{BI} \q
 \ea
 into two matrices.  $U^C_{AI}$ and $V^C_{BI}$ are the (unitary) matrices built from the singular vectors of $M^C_{AB}$. $\lambda^C_I \geq 0$ are the singular values of $M^C_{AB}$ and are ordered in decreasing order. This transformation is exact, but $M^C_{AB}$ can be approximated by only taking the largest $n$ singular values. Indeed this is the best approximation of $M^C_{AB}$ by a matrix of rank $n$ (with respect to the least square error).
 
Cutting off the singular values at the highest one, $\lambda_1$, we obtain a factorizing amplitude $M^C_{AB} \sim (S^{1})^C_A (S^{2})^C_B$. The $S^i$--matrices define the amplitudes for the prisms, e.g. $(S^{1})^C_A = A_{P_A} (\{k_e\}_{e \in {\mathbb A}},  \{k_e\}_{e \in {\mathbb C}})$.
 
Note that here the shared variables $\{k_e\}_{e \in {\mathbb C}}$ serve as parameters in the SVD procedure. In this lowest order approximation we do keep one singular value per index combination $\{k_e\}_{e \in {\mathbb C}}$, that is $K^3$ singular values in each splitting. 
 
 %We thus would summarize the edge labels of one prism to the index $A$ and the edge labels of the second prism to $B$.
 %But it seems that we need also to split four of the edges, which are shared by the two prism halves. Again because of the Gauss constraints (now for the new cut faces), these are actually just three independent variables.
 
 %We could copy each of the independent labels $k_e$ to a pair $k_e,k_{e'}$, multiply the cube amplitude with Kronecker delta, $\delta(k_e,k_{e'})$ and then proceed with a singular value decomposition into two matrices, where $k_e$ and $k_{e'}$ will be associated to the two different prisms.
 
 %However, the Kronecker delta $\delta(k_e,k_{e'})$ leads to matrix $M_{AB}$ in block form, where the blocks are labelled by the three independent shared edge variables $k_e=k_{e'}$. Thus instead of proceeding with an SVD for the full matrix at once (and just keeping one of the SV's) we perform a singular value decomposition for each of the blocks and keep one SV per block (thus $K^3$ SV's in total). 
 
 %That is, we actually perform SVD's of matrices $M^{\{k_e\}}_{A'B'}$ which are labelled by  three independent shared edge labels $\{k_e\}$.  The index $A'$ summarizes all the other (initially two independent) edge indices of one prism, and $B'$ the other (initially two independent) edge indices of the other prism. 

 This procedure is used to determine the amplitudes associated to the prisms $A_{P_i}$, with four possible prisms $P_i$ resulting from cutting the cube along the two possible face diagonals as depicted in figure \ref{fig:3d-splitting}.  The prisms are then glued to new (larger) cubes as depicted in figure \ref{fig:3d-splitting}. \\
 
\begin{figure}
\begin{center}
\includegraphics[width=0.3\textwidth]{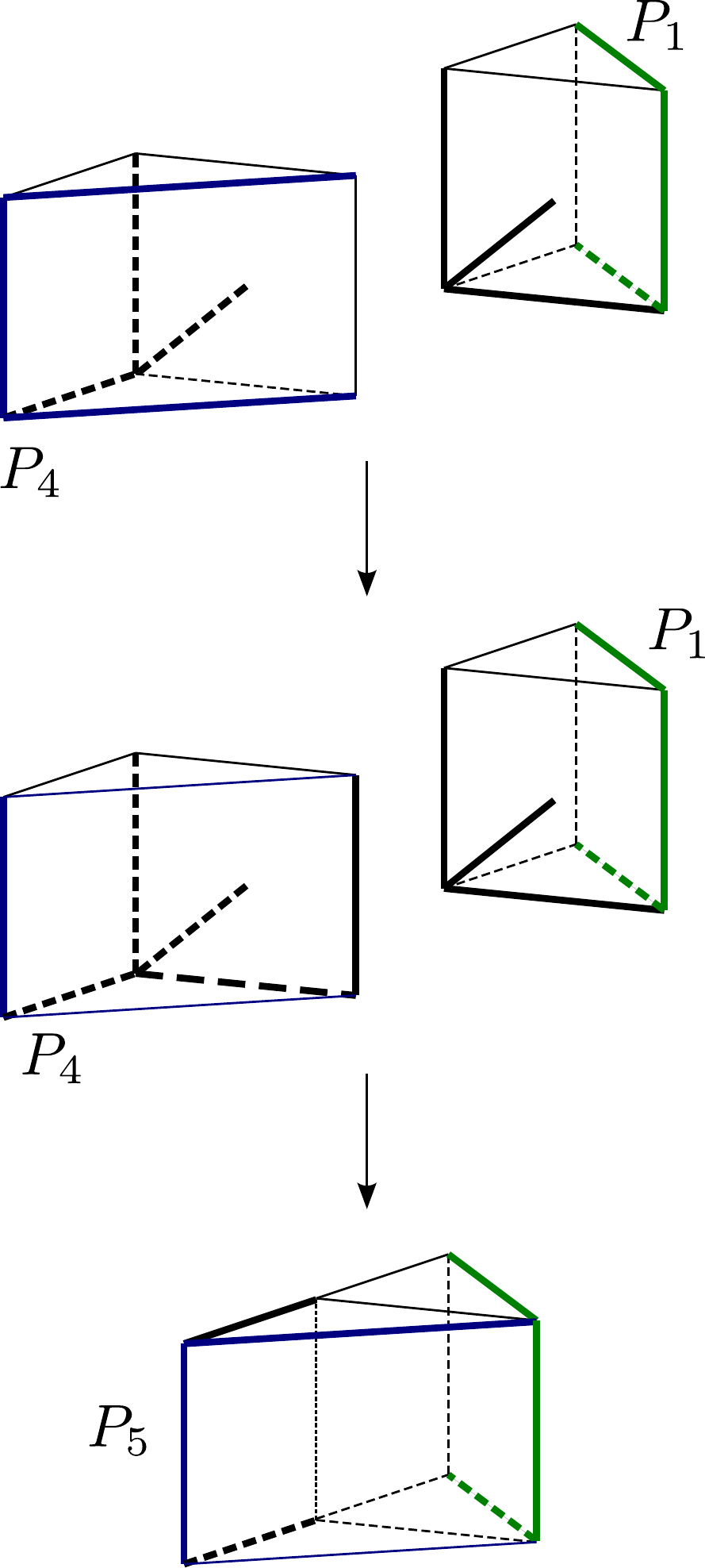}
\caption{Gluing of two prisms  into a larger prisms. This gluing requires a redefinition of the tree (represented by bold edges) for one of the prisms $P_4$.  After the gluing a further redefinition of the tree has been performed -- the final tree is shown in the bottom panel.%%For details refer to the text.
\label{fig:3d-glueing}}
\end{center}
\end{figure}

{\bf Gluing of the prisms into cubes:}
The prisms are glued in two steps to cubes: In a first step one glues $P_1$ with $P_4$ to a new (larger) prism $P_5$ and $P_2$ with $P_3$ to a new larger prism $P_6$. The gluing of these pairs is along the subdivided face with one additional (independent) edge -- a half--diagonal of this face. This edge carries the only variable $k_d$ that is summed over.  That is we divide again the variables of each $P_i$ into three sets: $\{k_e\}_{e \in \mathbb{S}} \equiv k_d$ is the (independent) variable that is summed over. $\{k_e\}_{e\in \mathbb{C}}$ are the independent variables associated to the boundary edges of the glued face. If necessary we redefine the tree\footnote{This is a simple linear transformation on the variables determined by solving the Gauss constraints for the different sets of independent variables as defined by the trees.} of the prism in question, so that the number of these independent variables is three. The edges carrying the three variables need of course to match with the ones of the  other prism, see figure \ref{fig:3d-glueing}.

The third set $\{k_e\}_{e\in \mathbb{R}_i}$ are the remaining (two) independent variables. The variables $\{k_e\}_{e\in \mathbb{C}}$ from the two prisms to be glued need to be identified with each other. Thus the amplitude for e.g. the prism $P_5$ is given by
\ba
&&A_{P_5} ( \{k_e\}_{e\in \mathbb{R}_1},\{k_e\}_{e\in \mathbb{R}_3},\{k_e\}_{e\in \mathbb{C}} )\nn\\
&=& \sum_{\{k_e\}_{e \in \mathbb{S}} } A_{P_1} (\{k_e\}_{e\in \mathbb{S}}, \{k_e\}_{e\in \mathbb{C}}, \{k_e\}_{e\in \mathbb{R}_1}) \nn\\
&&\q\q\q\q  A_{P_4} (\{k_e\}_{e\in \mathbb{S}}, \{k_e\}_{e\in \mathbb{C}}, \{k_e\}_{e\in \mathbb{R}_4}) \q .
\ea

In the second step one glues $P_5$ with $P_6$ to a new cube in a similar way as for the previous gluing.  The pair of prism faces which are now identified in the gluing are actually  subdivided into two smaller faces each by one `inner' edge.
However the variable $k_e$ associated to this edge is completely determined by the boundary edges via the Gauss constraints. Thus we do not have a variable to sum over in the gluing.  The to be identified prism face has six boundary edges which carry 5 independent variables of type ${\mathbb C}$. (Again we need to redefine the tree so that the edges of this tree agree with the edges carrying the independent variables.) Additionally there are two 
variables for each of the prisms $P_5$ and $P_6$ of type ${\mathbb R}$. As there is no summation involved the amplitude for the new cube is just given by the product of the two prism amplitudes. This new cube amplitude depends on nine variables, as did the initial cube amplitude (after introducing the subdivision of two faces into four triangles).

~\\

 This completes one coarse graining step in a plane. We now rotate the cubes, to coarse grain in the plane orthogonal to the previous one and start the process again with the splitting  of the cube into prisms. %It turns out that we only have to subdivide undecorated faces. During the gluing of the prism the only summation is actually over the decorated faces - thus for gluing a cube we just have to sum over two variables, as the edge shared by all four prisms is Gauss constrained by outer edges. 
 
 The most time consuming step is the last gluing step, which scales with the number of variables involved, that is with $K^9$ (with $K$ denoting the size of the gauge group). It  takes negligible time on a laptop for the Ising gauge model in this lowest approximation (less than a second for 50 iterations, which corresponds to coarse graining roughly 16 3D cubes).   By comparison, if we just consider the contraction of one index between two six--valent tensors (in a cubical 3D tensor network), it has computational complexity $\sim K^{11}$, with $K$ being the bond dimension.

The Ising gauge models flows either to the high temperature / strong coupling phase with $\lambda_1^{ \{k_1, k_2, k_3\} } = \delta_{k_1,1} \delta_{k_2,1} \delta_{k_3,1}$ or the low temperature / weak coupling phase with $\lambda_1^{ \{k_1, k_2, k_3\} } = 1\; \forall \, k_1, k_2, k_3$. 
In this lowest order approximation we observe the  phase transition at $\beta_c = 0.703$, within $8\%$ of the critical temperature $\beta_c = 0.761$ found by Monte Carlo methods \cite{montecarlo1,montecarlo2,montecarlo3}. By comparison, the simplest isotropic Migdal Kadanoff recursion relation \cite{MK1,MK2,kadanoff} leads to a phase transition $\beta_c = 0.683$.

%which is already the most accurate result obtained using tensor network renormalization techniques so far. Low computational complexity allows for future extensions into models with larger structure groups.

Apart from the simplicity of this algorithm compared to the other options, let us point out another advantage: this is in performing the SVD in blocks labelled by parameters. These parameters are given by the ${\mathbb C}$ type variables, which are straightforward to interpret. In fact, for spin foams the representation labels carry geometric information and represent in 3D the length of the edges. The same strategy applied to the spin net models provided invaluable information about the phase space structure \cite{qgroup} and in general offers an easy way to monitor the outcome of the coarse graining algorithm. For non-Abelian models, this SVD procedure is in parallel to the block form used in spin net models \cite{eckert1,eckert2,merce} and the block labels can be interpreted as intertwiner channels. This enforces the hope that spin nets and spin foams do indeed share statistical properties.

\subsection{Improving the truncation: introducing the tensor network}

 In order to improve the truncation one would take into account more than one singular value per block in the process of splitting of cubes into prisms\footnote{This can be done by fixing the number of singular values per block, and thus working with tensors or pre--determined size. A more complicated version \cite{eckert1,eckert2} is to compare singular values across blocks and introduce a global cut--off for the number of singular values. Thus the size of the blocks might actually vary. Experience with the models in \cite{eckert1,eckert2} has shown that reducing the size of a block might lead to instabilities, which can be addressed by introducing an algorithm in which the block size can increase, but not decrease.}. This requires the introduction of additional indices on the faces, here denoted by $I_f \ldots$, that describe the correlations between the two split parts of the cube. 
 
 These indices can be organized into a tensor network, where the network is situated on the dual lattice (i.e. vertices are placed in the middle of the cubes and the network edges cut through the faces of the cubes.) A key step of the algorithm is then again the splitting of the cube together with the tensor network vertex into two cubes with associated vertices. There are essentially two different ways to deal with this, which we will describe in the following.
 
 {\bf Higher order algorithm with additional face and edge indices:} After one has gone with the initial amplitude through one iteration, we will have additional indices $I_f$ on four faces of the cube. In the following iterations, faces are split into triangles and hence the face indices  have also have to be `divided': 
 In this version of the algorithm we do so by putting the face index $I_f$ onto the newly introduced diagonal edge $e$ and denote it by $I_e\,(=I_f)$. The diagonal $e$ now carries an index pair $(k_e,I_e)$.  Thus if we glue the entire complex, the index on the diagonal edge will ensure that the four triangular faces that are glued in pairs onto each other, carry the same index $I_e$, as is the case before cutting. Eventually, all the (square) faces will carry variables $I_f$, and all the edges (including half diagonals) will have  variable pairs $(k_e,I_e)$. The $k_e$ variables will still be restricted by the Gau\ss~constraints, whereas the $I_f$ and $I_e$ indices are unrestricted.  These additional indices can be understood as higher-order corrections decorating the cube amplitude, cf. figure \ref{fig:decorated}.

 %We define the initial amplitude as
%\ba\label{initialhigher}
%A_c =\prod_{e \in c} \left( \tilde\omega(k_e) \right)^{1/4} \prod_{e \in c} \delta_{I_e,1} \prod_{f \in c} \delta_{I_f,1} \; ,
%\ea
%where the introduced indices can be understood as (initially empty) higher-order corrections decorating the cube amplitude, cf. figure \ref{fig:decorated}.

Let us return to the process of splitting the cube into prisms, which is implemented by a SVD. There are now three independent indices $k_e$ and four indices $I_e$ that are shared by both prisms. These indices could therefore be used as parameters for the SVD, as described before in the case of just having the three $k_e$--indices. One would now have $K^3 \times \chi_{e}^4$ blocks where $\chi_e$ is the lengths of the $I_e$ indices. Assuming we choose also $\chi_e$ singular values per block we would take into account $K^3 \times \chi_e^5$ singular values in this splitting and thus have an `effective' bond dimension of $\chi_{\mathrm{eff}}=K^3 \times \chi_e^5$.

However we have now a situation where one expects that the singular values from a block labelled by $I_e=2$ for all four edges are much smaller than from the block labelled by $I_e=1$. This can be accommodated by comparing the singular values for all block labels among each other and selecting the $\tilde \chi$ largest ones, where $\tilde \chi$ is now a choice for the effective bond dimension. The $\tilde \chi$ largest singular values have to be distributed among the various blocks, resulting into blocks of varying size and also into  vanishing blocks if these do not get any singular values assigned.\footnote{See \cite{eckert1,eckert2,eckertdiploma} for an implementation of an algorithm with varying block size in 2D} The size of an index $I_e$ can thus  get reduced  a posteriori. 
This makes the algorithm much more complicated to implement. A possibility, to be explored in future work, is to fill entries up with zeros to reach a constant block size and then dealing with sparse matrices or  tensors.

An alternative procedure is to keep just the block form with the $k_e$ indices as parameters and to accommodate the $I_e$ indices by first multiplying the amplitude with Kronecker delta's $\delta_{I_eI'_e}$ and then to split by an SVD with the $I_e$ indices assigned to one prism and the $I'_e$ indices assigned to another prism. This automatically compares the singular values across the $I_e$ block structure, with a constant block size $\chi_e$ (i.e.\ $K^3 \times \chi_e$ singular values are taken into account with each splitting) , however with the disadvantage that one has to split factors of Kronecker deltas with an SVD. 
 
 After the splitting of the cubes into prisms one proceeds as before, gluing four prisms to a cube in two steps. We sum over only the bulk variables, i.e.  $I_e$ and  $k_e$ labels on the half diagonals, as well as face indices $I_f$ for the faces without half diagonals. The most expensive step is the gluing of the two pairs of prisms into the cube, the cost of which scales with $K^9 \times \chi_e^{27}$.

A reduction in costs can be obtained by summarising the variables on the top and bottom faces (with total bond dimension $\chi^4_e$ each), $T$ and $B$ respectively, into single face indices, $T'$ and $B'$ (treating the one independent $k_e$ index of one half diagonal as a parameter). Such a truncation map, introducing a cut-off $\chi^4_e \rightarrow \chi_e$, can be implemented using the SVD:
\ba\label{embeddingmap}
A_{ \{ BRT \} } =\sum_{{T'}=1}^\chi U_{ \{BR\},{T'}} \lambda_{T'} V^*_{ {T'}T } \, ,
\ea
where $R$ denotes all the remaining indices. After applying the truncation map on both sides   (here $(V^\dagger)_{TT'}$ from the right and $V_{B'B}$ from the left) the new truncated amplitude becomes 
\ba
A_{ \{ B'RT' \} } =\sum_{B=1}^{\chi^4} V_{B'B}  U_{ \{BR\}, T'}\lambda_{T'} \, .
\ea
In fact we could have also chosen $U$ as a truncation map. We can apply the technique from \cite{beijing} to determine which of the maps, $U$ or $V$, gives a better truncation by comparing the root mean square values of the rejected singular values in both cases. 

The simplification reduces the memory usage from $O(K^9 \times \chi_e^{24})$ to $O(K^9 \times \chi_e^{18})$,  which allows for numerical testing of this higher-order algorithm. For the 3D Ising gauge model we obtain an improved critical temperature, $\beta_c = 0.710$ for $\chi_e=2$. This is quite a moderate increase in accuracy given that the increase in computational costs are substantial.

 The memory and computational costs of this improvement scheme are quite high, also because additional indices are associated to faces {\it and} edges. An alternative scheme which we will discuss now only introduces additional face indices which will reduce the costs significantly.

{\bf Higher order algorithm  with additional face indices only:}

 Imagine we put a vertex in the middle of the cube and attach a tensor with six edges carrying the $I_f$-type indices. Then the splitting of a cube would entail the splitting of a tensor network edge. To do so we proceed as follows: We formally replace the face index $I_f$ with a pair $(I_f,I'_f)$  (for the two triangular faces) and multiply the amplitude of the cube with two Kronecker deltas $\delta(I,I')$ for the two pairs of triangular faces. We then proceed with an SVD, treating the $k_e$ indices as parameters but not the two $I_f$ indices that are being split. Thus, the number of singular values taken into account at each splitting is $K^3 \times \chi_f$, that is $\chi_f$ singular values per block labelled by the three independent indices $k_e$ of the newly introduced face. This defines the effective bond dimension, i.e.\ the  number of indices per tensor leg, as $K^3 \times \chi_f$.

Proceeding as before with gluing the prisms to cubes, the cost of the algorithm goes as $K^9 \times \chi_f^{14}$. We can reduce it further by applying the embedding maps on the top and bottom faces in two steps: $\chi^2_f \rightarrow \chi_f$ after gluing two prisms, and once again after two pairs of them are put together. The overall complexity of the algorithm scales with $K^9 \times \chi^{10}_f$. By comparison the (higher order singular value) 3D algorithm of \cite{beijing} has scaling of $\chi^{11}$ for the computational costs (for models that can be brought into tensor form with bond dimension $\chi$). For the Ising model we found a steady improvement of the critical temperature with increasing $\chi_f$, reaching $\beta_c = 0.729$ for $\chi_f=8$.

 ~\\
This and the previous kind of algorithm can be seen as using cubes with a more complicated subdivision of faces. Choose for instance $\chi_f=K$. This exactly corresponds to the number of  additional variables which we obtain by subdividing each face by its two diagonals into four triangles. The four half diagonals are constrained by three independent Gau\ss~constraints, hence we have one additional variable $I_f=k_f$ per face. The copying of the face index and the associated introduction of Kronecker deltas $\delta(I,I')$ corresponds to the fact that the Gau\ss~constraints leave only one independent additional variable $k_f$, although there are four half diagonals. 

In this interpretation the algorithm that introduces additional face and edge indices and the algorithm that introduces only face indices differ in how they treat the Gau\ss~constraints resulting from the triangular faces:  In the first case one shifts (using the Gau\ss~ constraints) the $k_f$ index onto the diagonal along which the face is subdivided. This diagonal is now split into two (half-) diagonals, where one of the half diagonals carries  the additional index. (The variable for the other half diagonal is determined by the two outer edges and the Gau\ss~constraints.) The Gauss~constraints are exactly preserved also for the finer variables. In the other case we rather shift the $k_f$ index onto one of the half diagonals not coinciding with the cut.  The variable for the other half of the diagonal is again determined by the Gau\ss~constraints as long as one has not split the square face into two triangular ones.  The other independent variables are given by the three outer edges of the square face, which also determine the variable associated to the (entire) cut diagonal. 
One now proceeds with the splitting of the cube where the finer Gau\ss~constraints are taken as part of the amplitude instead of being preserved explicitly.

\subsection{Algorithm with lower computational costs}

The 3D algorithm still requires large computational resources, especially related to memory consumption of the decorated tensor, which largely pertains to dealing with the dimension higher than two. However, another reason is that the algorithm is based on cubes, which are not the most elementary building blocks. A 3D cube has 12 edges and 6 faces, whereas a tetrahedron has only 6 edges and 4 faces. The difficulty is to come up with regular coarse graining schemes involving tetrahedra. One straightforward scheme of coarse graining tetrahedra into tetrahedra right away is based on cubes subdivided into 6 tetrahedra each.   To do so, we first divide the cube into two parts as in our algorithm, along the diagonal of one of its squares. These halves get triangulated by three tetrahedra as follows: One tetrahedron contains the upper triangle of the half-cube and is completed by connecting it to the vertex beneath the tip of the triangle. Thus all other remaining squares are also cut in half. For the second tetrahedron, you take one of these triangles and connect all vertices to another vertex such that the main diagonal of the cube is an edge of the tetrahedron. The remaining part is the third tetrahedron. The second half-cube is cut analogously to match the edges of the first one. In this case 8 smaller tetrahedra are coarse grained into a larger tetrahedron with all edges doubled in length. 

\begin{figure}
\begin{center}
\includegraphics[width=0.45\textwidth]{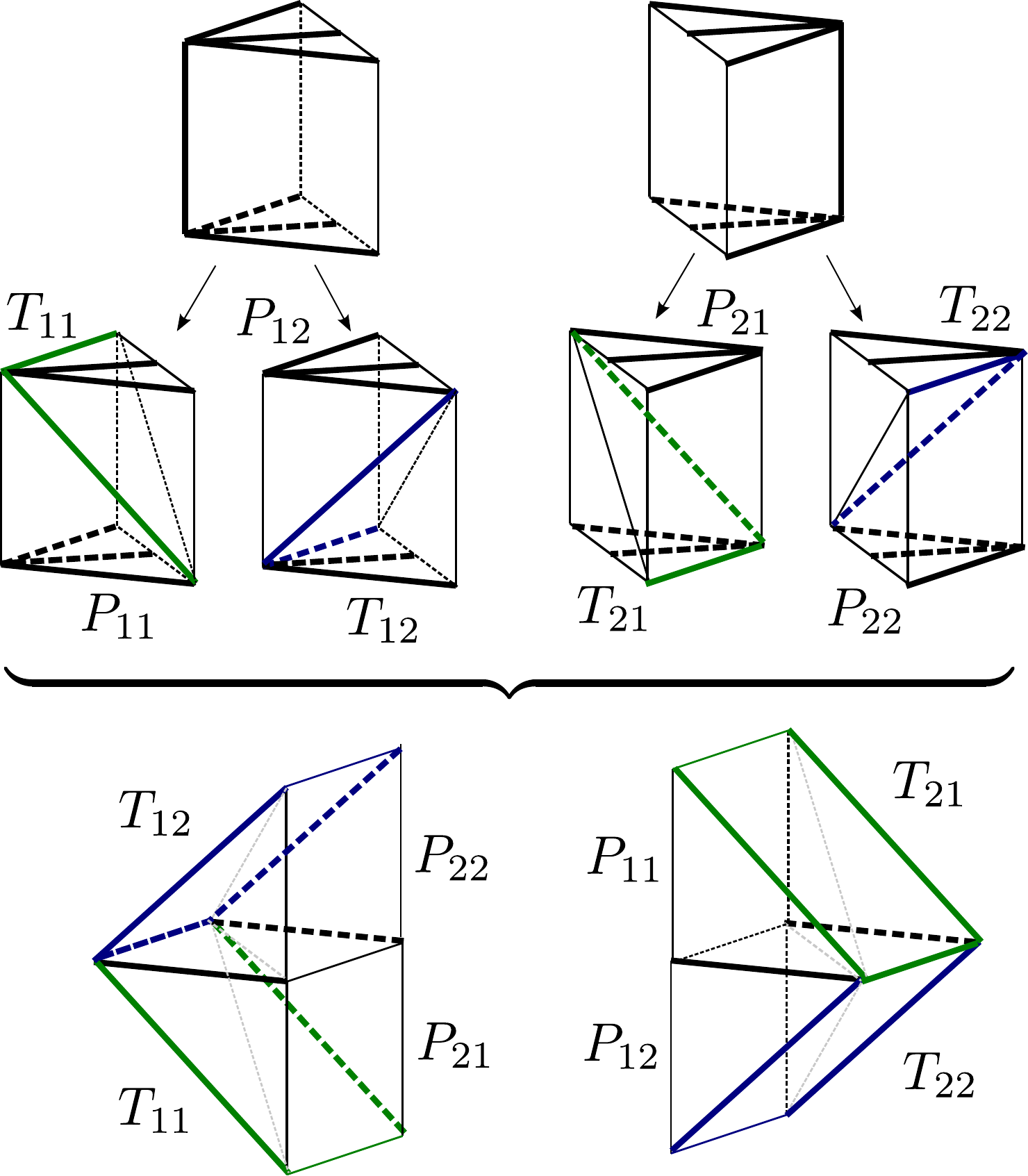}
\caption{Simplified algorithm based on prisms, tetrahedra and pyramids. Basic building blocks are now prisms, which after splitting and re-glueing form the coarse-grained polyhedra.
\label{fig:ptp-algo}}
\end{center}
\end{figure}

Here we propose an algorithm based on prisms, which can be obtained from the cube algorithm described above by cutting the cubes. Instead of starting with cubes and subdividing these into prisms, we start with prisms right away, see figure \ref{fig:ptp-algo}. We divide two of these into pairs of tetrahedra and pyramids, before re-glueing into larger prisms back again.

The advantage is, that in the leading order approximation, the basic amplitude carries fewer indices, decreasing the computation cost to $O(K^8)$, as well as the memory consumption from $O(K^9)$ to $O(K^7)$. While we find no qualitative differences to the cubic algorithm, the obtained critical temperature is near $\beta_c = 0.630$ (within $18\%$ of the Monte Carlo result \cite{montecarlo1,montecarlo2,montecarlo3}). Although less accurate, the technique might find applications in the models with large structure groups, where the cubic algorithm would be too expensive.

\section{Outlook: non-Abelian models and four dimensional models}

The technique is in principle also applicable to lattice gauge theories (and spin foams) with non--Abelian structure groups. The details are however much more involved than in the Abelian case and will therefore appear elsewhere, here we will only provide a sketch for the lowest order approximation. A discussion of the different representations (and tensor network renderings)  of lattice gauge theory and spin foams can be found in the appendix.

First of all the amplitudes for the cubes can be organized in a similar way as for the Abelian case. We now  must however associate also representations to the diagonals of the cubes. Additionally the representations for non--Abelian groups are in general larger than one--dimensional. This results in an additional pair of magnetic indices that have also be associated to the edges. Each edge and diagonal carries hence indices $(\rho_e,m_e,n_e)$ where the number of index values  for $(\rho_e,m_e,n_e)$ coincides with the size $|G|$ of the non--Abelian  (finite) group $G$. 

The amplitudes carry however a notion of gauge invariance -- which as before reduces the number of truly independent variables. There are two different ways to use this gauge invariance:

{\bf By gauge fixing the cube amplitudes:}  We have a priori indices $(\rho_e,m_e,n_e)$  associated to each of the 12 edges and 6 diagonals of the cube. Gauge fixing  along a tree (which in the Abelian case amounted to solving the Gau\ss~constraints and can therefore here been seen as a non--Abelian equivalent) leads only to 7 independent variable triples $(\rho_e,m_e,n_e)$.  As in the Abelian case we obtain two additional independent triples corresponding to the subdivision of two faces into four triangles each. Thus we have 9 indices with bond dimension $|G|$ -- which coincides with the Abelian case.

Note that this reduction leaves one global symmetry (by adjoint action of the group) for the cube. This symmetry is trivial in the Abelian case. 

The algorithm proceeds as in the Abelian case by splitting the cubes into prism and gluing the prisms into larger cubes. The behaviour of the various kinds of indices under splitting and gluing is however more involved and will be detailed elsewhere. A further step that is required is a change of the tree defining the gauge fixing  -- as in the Abelian case a transformation of the tree determining the independent variables. However in  contrast to the Abelian case the transformation is now more involved with costs $\sim O(|G|^{10})$. This is one determining factor for the costs of this algorithm, the other being the last gluing step which has costs bounded by $O(|G|^9 \times D^5)$ where $D$ denotes the largest dimension among the unitary irreducible representations of the group $G$.

{\bf Fully symmetry protected version:} Alternatively one can parametrize the amplitudes in a fully gauge invariant manner.  This can be actually understood as a unitary transformation that acts on the magnetic indices in the previous parametrization and converts these to intertwiner labels, which for multiplicity free groups\footnote{ For a multiplicity free group the tensor product of three representation contains maximally once the trivial representation as a summand.} can themselves be taken as representation labels (in a recoupling scheme).  If the transformation converts all the magnetic indices into representation labels the transformation is unitary if considered  from the subspace of amplitudes invariant under the global adjoint action. This form of the amplitudes is known as the `spin representation' for spin foams and appears for the strong coupling expansions for lattice gauge theories.

This representation leads to 18 variables $\rho$ attached to the edges and diagonals of the cube. The $\rho$ labels attached to the diagonals,  can be understood as indices associated to the faces. However there are complicated coupling rules (generalizing the Gau\ss~constraints for the Abelian case) for the representation labels -- thus the allowed range of the 18 variables is much smaller than  (number of irreducible unitary representations) to the power of 18. In fact the number of index combinations is between $|G|^8$ and $|G|^9$, as it should agree with the number in the previous parametrization and in addition take the global adjoint action\footnote{The orbits of this action can vary in size from one point to $|G|$ points.} of the structure group into account.  

This can be dealt with by organizing the representation labels into super indices (see for instance \cite{qgroup}). The precise form of these super indices has to be adjusted to each splitting  step and gluing step in the algorithm.  
~\\ 

Let us also point out that the algorithms can be generalized to four dimensions. We have now hypercubes, the (in 3D) edge indices $\rho_e$ are replaced by plaquette indices (i.e. attached to the two--dimensional objects) $\rho_p$ and the (in 3D) face indices $\rho_f$ are now attached to the 3D cubes which make up the boundary of the hypercube. Thus the splitting and coarse graining can as before be considered to take place in a plane.

The memory costs for one hypercubical building blocks (at lowest order approximation) scales with $|G|^{17}$. As in 3D  we thus expect the computational time cost to be of roughly the same order. This memory cost has to be compared with the costs for the tensor network representation discussed at the end of the second section: for a hypercubical lattice and an Abelian group this is given by $|G|^{40}$ following from the fact that one has tensors with eight indices, each of which has bond dimensions $|G|^5$. The computational time costs for gluing two such tensors along one edge is hence $|G|^{75}$.

Thus the difference in costs between a decorated tensor network procedure and a pure tensor network procedure are very high -- but the absolute costs of the decorated tensor network procedure are nevertheless quite large, although doable for small groups -- the most expensive algorithm we tested in 3D had memory costs $\chi^{27}$. For this reason it might be worthwhile to shift to more elementary building blocks, as we have done in the 3D tetrahedral-prism algorithm. For instance the memory costs for a 4--simplex scales just with  $|G|^6$. By comparison the memory cost of a cubical building block and a tetrahedra in 3D are $|G|^7$ and $|G|^3$ respectively. Thus the potential savings of shifting to simpler building blocks in 4D are much higher than in 3D.

\section{Decorated tensor networks for other models}

 In this section we lay out that the decorated tensor network algorithm can also be applied to other  models, besides (generalized) lattice gauge theories.
 
Let us describe the general structure of decorated tensor models for lattice gauge theories: A tensor carrying the indices $I,J,M, \dots$%, that are trivial for the lowest order approximation,
sits in the centre of each cube with its edges piercing the faces of the cube, equipping them with indices. The tensors in neighbouring cubes are glued together by contracting the index of the edge connecting them, which is equivalent to summing over the index on the common face. In addition, this tensor network comes with supplemental decorations, i.e. the labels attached to the edges of the cube. In fact these are the variables which keep most of the physical information. 
 Indeed one can imagine to contract the `internal' tensor network by summing over all indices $I,J,M,\dots$ and be left with a model with the original type of variables.

 This model would in general be non-local, if the $I,J,M,\dots$ indices have non--trivial range. Thus we can directly understand how non--local couplings, which appear in other real space  renormalization schemes,  are turned into additional indices and  how the cut--off on these indices can also be understood as a truncation of non--local interactions that can occur, see also the discussion in \cite{dittcyl,timeevol}.

\begin{figure}
\begin{center}
\includegraphics[width=0.455\textwidth]{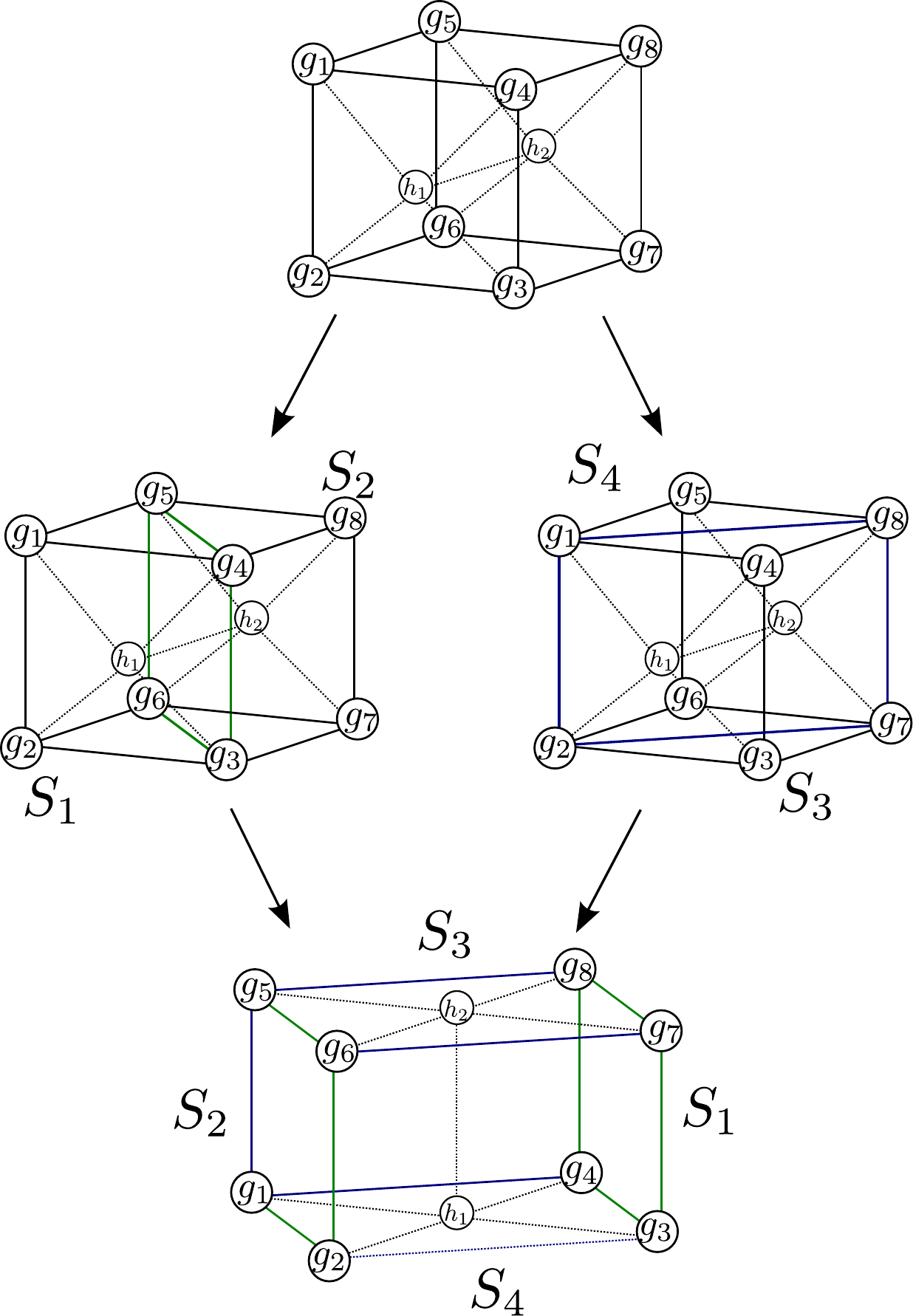}
\caption{Schematic illustration of the 3D algorithm: Similar to the gauge theory version, cubes are split along their diagonal faces, here labelled by four spins that are split into two. As a result, one considers generalized cubic amplitudes, where the `crossed' faces carry an additional Ising spin.
\label{fig:3d-ising-algo}}
\end{center}
\end{figure}

Let us proceed with the application of this idea to Ising--like models in 2D\footnote{A similar algorithm as just presented for the 3D gauge models can also be applied to 3D Ising--like models, see also figure \ref{fig:3d-ising-algo}.} on a square lattice. The algorithm will amount to a sophisticated decimation procedure, i.e. keep a subset of the original Ising spins, separated at larger and larger distances, while the decimated spins are absorbed into a tensor network.
%\footnote{Note that in a (flawless) decimation procedure the correlation function between two (non--decimated) spins does not change -- by definition. To compute the correlation function 
%%$\langle \sigma_i ; \sigma_j \rangle = \langle \sigma_i \sigma_j \rangle - \langle \sigma_i \rangle \langle \sigma_j \rangle$
% one has to consider all possible configurations of all spins; decimation only implies a preferred order of summation. Of course this argument is only valid as long as no approximations are applied in the decimation procedure. }
 These Ising spins can serve as signature observables, for instance one can straightforwardly compute a correlation function from the final tensor (this will be however at e.g. \ a distance half the system size in the contraction of  six tensors to a torus topology).

Again instead of starting with a tensor or a vertex model, we rather think of squares as the basic building blocks. To these squares we assign an initial amplitude, which is a function of the four Ising spins sitting on the four corners of the square. Additionally, we will use a dual lattice for the tensor network, i.e. each square is decorated with a vertex in its centre with four emanating edges that cut the edges of the squares at the midpoints. The indices on these tensor edges will allow to go to a higher order approximation. Thus we again obtain the structure of a decorated tensor network.

\begin{figure}
\begin{center}
\includegraphics[width=0.48\textwidth]{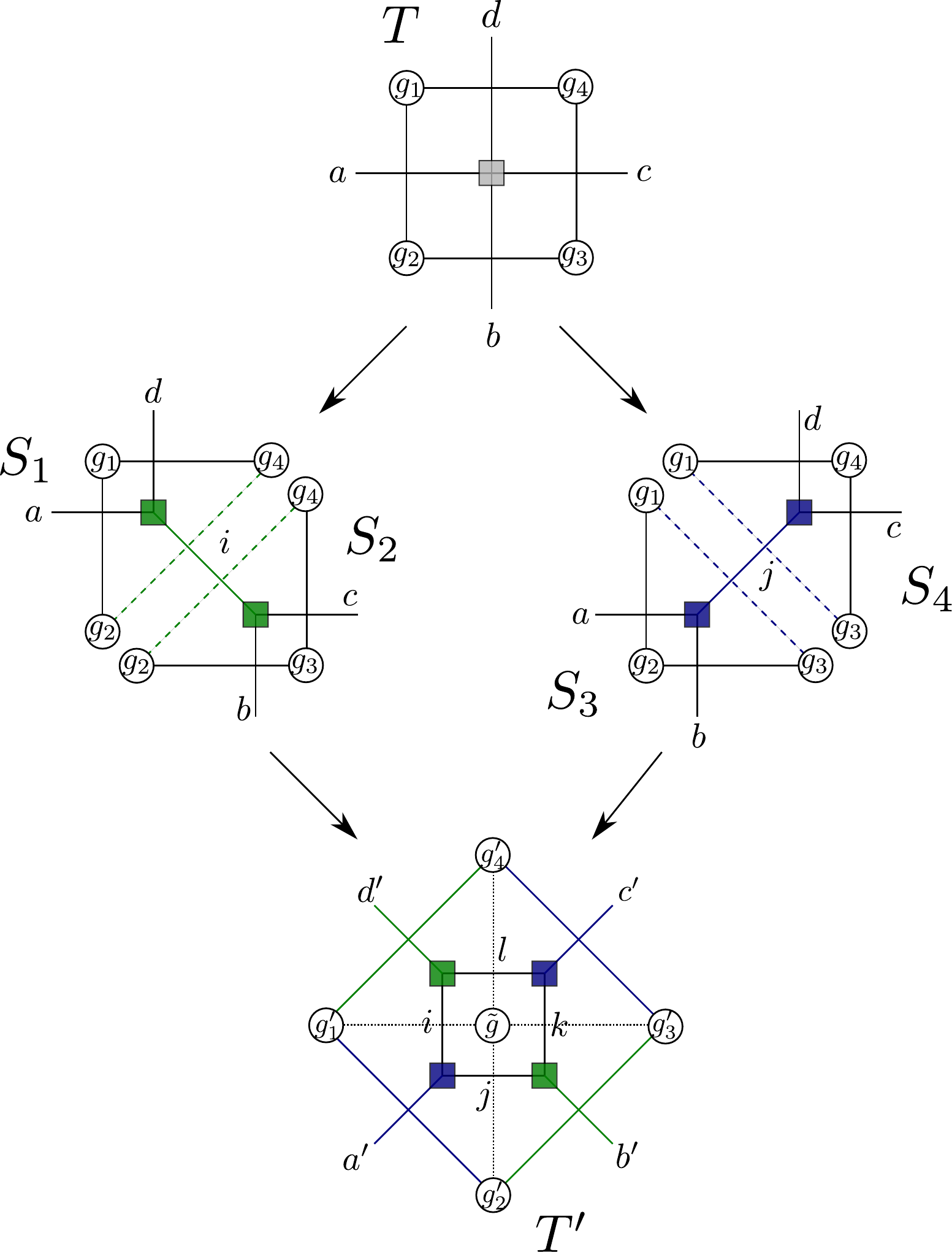}
\caption{ The initial amplitude $T$ associated to the square is split along its two diagonals. The two spins $(g_i,g_j)$ connected by the diagonal get duplicated and label the blocks to which SVD is applied. This gives four new triangular amplitudes, $S_1, \dots, S_4$, which are glued together to a new square amplitude $T'$. The last involves summation over the bulk tensor indices and one spin $\tilde{g}$.
\label{fig:2d-algo}}
\end{center}
\end{figure}

The coarse graining algorithm is a straightforward adaptation of the Gu--Wen algorithm \cite{gu-wen} taking the decoration by the Ising spins into account, see figure \ref{fig:2d-algo}: The idea is to cut the squares along their diagonals into two  triangles. One then glues four such triangles into a larger square. Thus we again have to discuss the splitting and the gluing procedure. 

{\bf Splitting:} As in the Gu--Wen algorithm, this requires to split the 4--valent tensor in the center of the square into two 3--valent ones, but additionally one has to `split' the Ising spins connected by the diagonal. 

Naively, one can double these spins and identify them by introducing Kronecker Deltas, which allows us to straightforwardly perform the SVD, yet with a significant amount of irrelevant data. Instead it is more efficient to use the pair of split spins as block labels: the matrix that we intend to split decomposes into smaller block matrices, labelled by the pair of spins, on which we apply a SVD separately. 

To be precise consider the splitting of the square in the top panel of \ref{fig:2d-algo} into two triangles as in the left centre panel.  In the notation of this panel we define the super indices $A=(g_1,a,d)$ and $B=(g_3,b,c)$ as well as $C=(g_2,g_4)$. This allows us to define a set of matrices $M^C_{AB}= A_{square}(a,b,c,d, \{g_i\}_{i=1^4})$ labelled by $C=(g_2,g_4)$. We perform a SVD for each of these indices so that we can approximate
\ba
M^C_{AB}= \sum_{i=1}^\chi (S_1)^C_{A,i} (S_2)^C_{B,i}  \q . 
\ea
This gives the amplitudes for the triangles (after converting the super indices back to the original indices).

{\bf Gluing:}
 After splitting the square along both diagonals, we obtain four triangular amplitudes which are connected to form the new (rotated) square as in figure \ref{fig:2d-algo}. This incorporates the summation over four tensor indices, as in the Gu--Wen algorithm, and one Ising spin in the center of the new square:
 \ba
&& A'_{square}(\{g'_i\},a',b',c',d')= \nn\\ &&\sum_{\tilde g, i,j,k,l}  A_{S_1}(j,k,b',g'_2,g'_3,\tilde g) A_{S_2}(i,l,d', g_1',g_4',\tilde g) \nn\\&&
 \q \q \; A_{S_3}(i,j,a',g_1',g'_2,\tilde g) A_{S_4}(j,k,b',g_2',g'_3,\tilde g)  \q .
 \ea
 
  In the lowest possible approximation, i.e. tensor index range one, one only sums over the Ising spin.

%The diagonals also cut through two of the Ising spins. In fact we should again do a SVD for each block, which are labeled by the values of these two Ising spins. After the splitting we have triangular building blocks, where the longer edges (or rather the edge crossing these longer edges) carry the  new tensor index. We glue four of these triangular pieces as in figure \ref{}. This involves a sum over the central Ising spin (which was not serving as a parameter in the previous SVD) and a summation over four of the tensor indices. Thus to the lowest possible approximation we just have to sum over one spin. In the end one ends up with (rotated) squares. 

\subsection{Simplification -- Triangular algorithm} \label{triangular}
 
 In fact, the code depicted in figure \ref{fig:2d-algo}, and also the related Gu--Wen algorithm, can be simplified by shifting the perspective from square building blocks to triangular ones. This simplified algorithm computes the same object (partition function for the square lattice) with same truncation as before, just that the costs will decrease. (This is different for the 3D tetrahedron - prism based algorithm, which is less costly than the cube - prism based algorithm but also decreases the quality of the truncation.)

If we consider the second part of the iteration, i.e. the glueing of the triangles and contraction of the respective indices, we realize that this can be performed in two steps: First we glue two small triangles into a larger triangle, then we glue the two larger triangles together to form the new square. However, in the next iteration of the algorithm we split this square again into the larger triangles we previously glued.

\begin{figure}
\begin{center}
\includegraphics[width=0.45\textwidth]{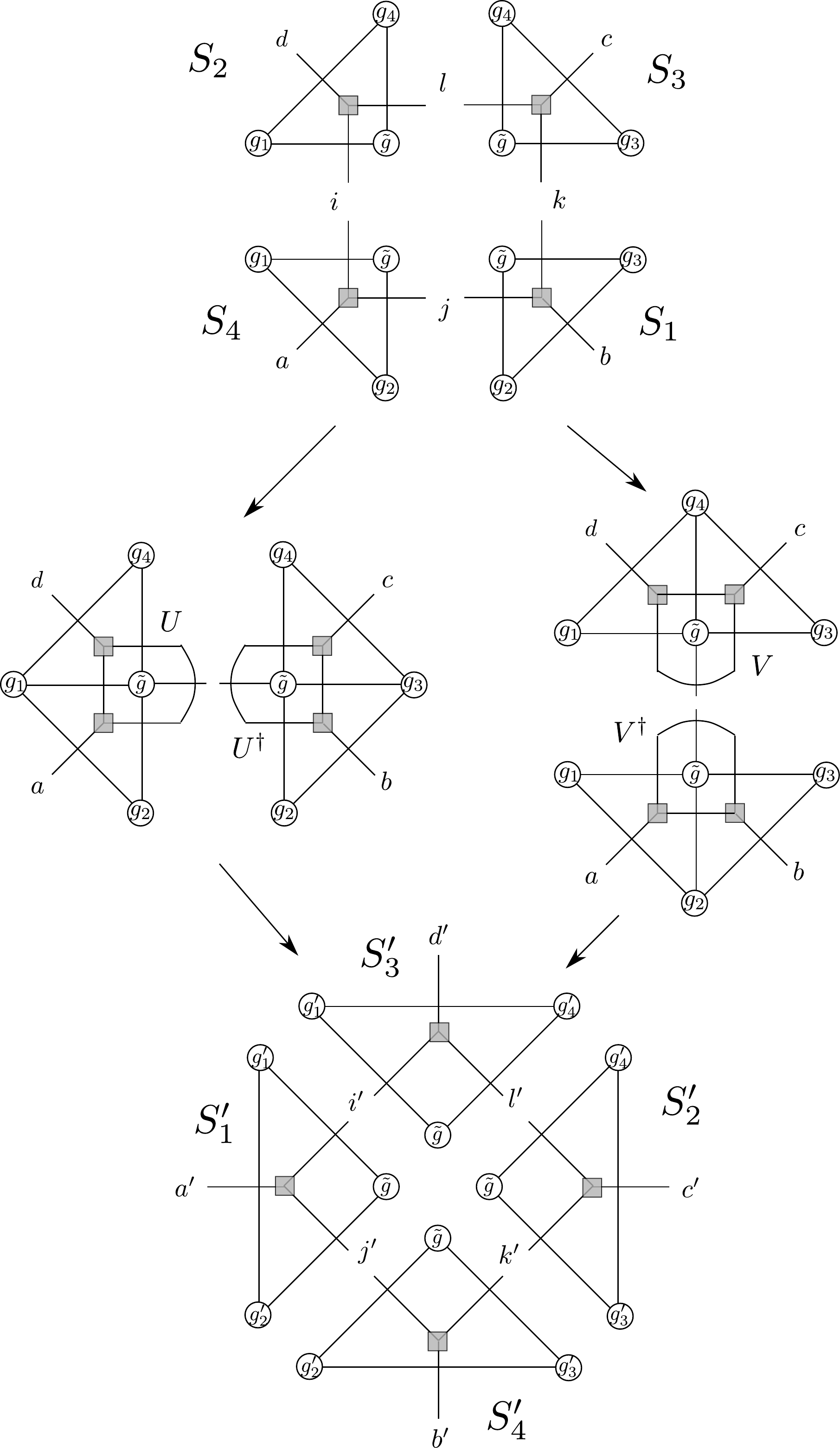}
\caption{  Triangular version: Starting from four triangular amplitudes, $S_1, \dots, S_4$, we glue them once vertically and once horizontally into larger triangles. The long edge carries two tensor indices and one spin, which we combine into a new tensor index by the embedding maps $U$, $U^\dagger$ and $V$, $V^\dagger$ respectively. These larger triangles, $S'_1, \dots, S'_4$, are the new triangular amplitudes for the next iteration.
\label{fig:2d-tria-algo}}
\end{center}
\end{figure}

Hence, instead of performing the summations and subsequent splitting of the square, we propose an algorithm purely based on triangular amplitudes, see figure \ref{fig:2d-tria-algo}. Therefore we keep track of four different triangles, denoted as $S_1,\dots,S_4$: these amplitudes are pairwise glued together, once vertically, once horizontally, forming new larger triangles, where the new long edge carries two tensor indices and one Ising spin. 

For instance glueing $S_2$ and $S_4$ in the top panel of \ref{fig:2d-tria-algo} amounts to:
\ba
&&A_{S_2-S_4}(d,a,j,l,g_1,g_2,\tilde g,g_4) = \nn\\
&& \sum_{i} A_{S_2}(i,l,d, \tilde g, g_4, g_1) A_{S_4}(j,i,a,\tilde g,g_1,g_2)  \q .
\ea

To arrive back at an amplitude comparable to the initial one, we have to define a truncation map (three--valent tensors) blocking the indices of the new larger side of the triangle into a new common tensor index, whose index range determines the quality of the approximation. Again these maps are computed via a (Higher Order) SVD as in  \cite{beijing}; note that the truncation maps applied to opposing triangles should be related by hermitian conjugation, such that they form a partition of unity (before implementing the truncation on the number of singular values).

That is for determining the truncation map for the triangle $S_2-S_4$ we introduce the following super indices: $A=(d,a,g_1)$, $B=(j,l,\tilde g)$ and $C=(g_2,g_4)$. We then form the matrices $M^C_{AB}= A_{S_2-S_4}(d,a,j,l,g_1,g_2,\tilde g,g_4)$ and perform an SVD for each $C$ configuration:
\ba
M_{A B}^C \, =\, \sum_m  W^C_{Am} \lambda^C_m U^C_{Bm} \q 
\ea
with $W^C,U^C$ unitary matrices.

From this SVD we only need $U^C_{Bm}$ where we restrict $m$ to run from $1$ to the bond dimension $\chi$. We apply this (now truncation) map to $M^C_{AB}$
\ba
(M')^C_{Am} = \sum_B  M^C_{AB} U^C_{Bm} \; ,
\ea
where $(M')^C_{Am}$ gives the amplitude of the new larger triangle $S'_2$ depending on variables $(A,C,m)$. This truncation map $U$ should then be also used for the truncation of $A_{S_3-S_1}$, which gives the amplitude of the larger triangle $S'_1$. One proceeds in the same way for the horizontally glued triangles in the left centre panel giving the amplitudes for the larger triangles $S_4'$ and $S'_3$.

This modification of the algorithm is not restricted to decorated tensor networks and can be readily applied also to the original Gu--Wen algorithm, for which it leads to a reduction of computational costs from $\chi^6$ to $\chi^5$.

% In the second step of each iteration we actually glue triangles that are split in the next step. The gluing can be performed in two parts: first two triangles are glued into a larger triangle and then these larger triangles into a square. As the square is split again into the larger triangles we do actually not need to perform this second summation. This results in an algorithm based on triangles, depicted in figure \ref{}. Such a reorganization of splitting and gluing is also possible for the original Gu-Wen algorithm for which it leads to a reduction of costs from $\chi^6$ to $\chi^5$.
 
The interesting feature of this decorated tensor network algorithm is the mixture of keeping part of the original variables and allowing for the additional tensor indices field redefinitions. An essential ingredient is a SVD per block  as extensively used in the algorithms developed  in \cite{eckert1,eckert2,merce,qgroup}. Especially for the non-Abelian models the knowledge of the singular values  per block was key to be able to interpret the fixed point tensors. This might help to  address some criticism to tensor network methods in \cite{kadanoff}, which points out the difficulty in extracting physical information from the fixed point tensors and separating physical from  unphysical  (due to the field redefinitions there is a kind of weak gauge symmetry for the tensor models) information. In terms of computational efficiency the algorithm is rather the same as the original Gu--Wen algorithm (apart from the decrease in costs due to basing the algorithm on triangles instead of squares).

%A rather inefficient version of this kind of algorithm would be to copy each of the variables on the corner of the squares to the neighbouring tensor edges (two for each corner) and then to impose by Kronecker deltas that the variables copied to adjacent edges are the same. This results in a tensor network of standard form, but now with a doubled initial bond dimension and an amplitude involving many Kronecker Deltas. This is the equivalent of using an algorithm based on some of the representations for lattice gauge theories discussed in section \ref{lgt}. 

~\\

\begin{figure}
\begin{center}

\includegraphics[width=0.45\textwidth]{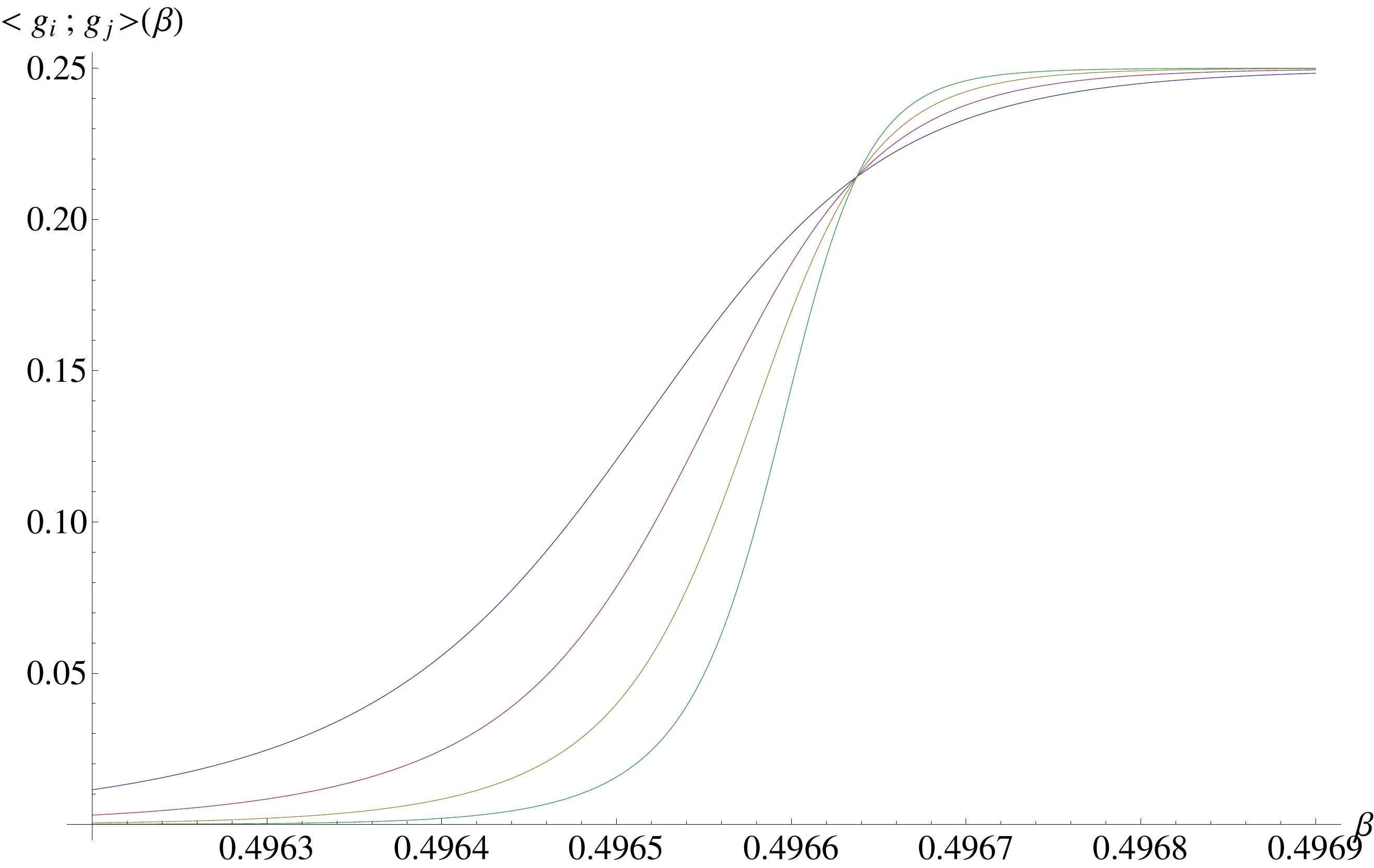}
\includegraphics[width=0.45\textwidth]{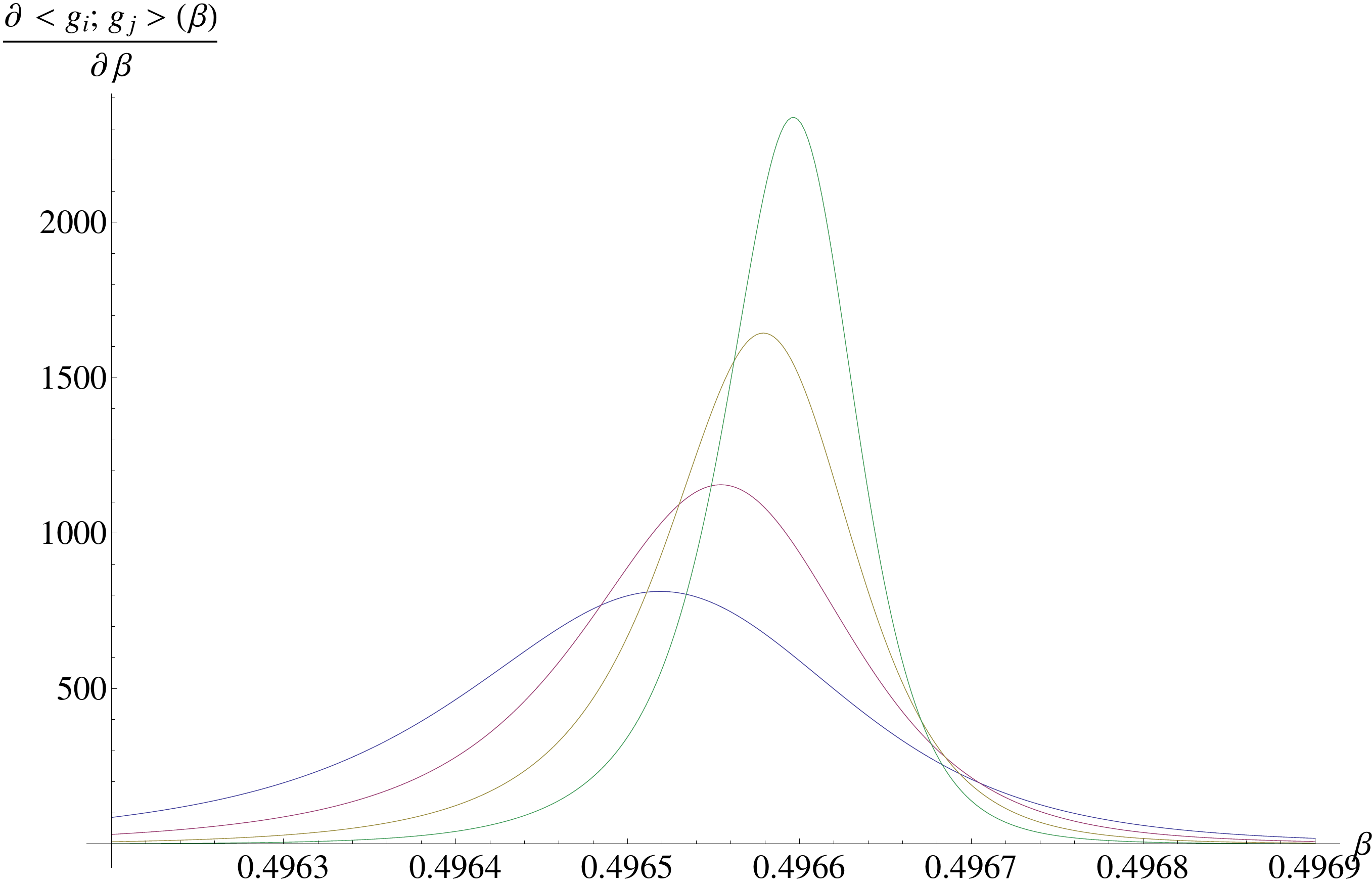}
\caption{  The correlation functions (top) and their first derivative (bottom) for two neighbouring Ising spins evaluated for amplitudes after different number of iterations, here $n=20,21,22,23$, in the lowest order approximation, $\chi_{\mathrm{eff}}=4$, with periodic boundary conditions. Due to the different number of iterations, the correlations functions correspond to different system sizes and different distances of the two spins on the original lattice. The more iterations have been performed the steeper the correlation function becomes on the phase transition and its first derivative diverges. The correlation functions for different number of iterations meet in one point, which marks the phase transition temperature.
\label{fig:corr-len}}
\end{center}
\end{figure}

 Results: We have applied decorated tensor network methods to the 2D Ising model defined on a square lattice. To lowest approximation,  i.e. $\chi=1$ per configuration of spins on the new edge corresponding to an effective bond dimension $\chi_{\mathrm{eff}}=4$,
 we observe a phase transition at the inverse temperature $\beta \approx 0.49664$. In comparison to the exact solution, $\beta_c = \frac{\ln (1 + \sqrt{2})}{2}\approx 0.44068$, this is an error of roughly 13 \%, yet still an improvement over the Migdal--Kadanoff methods, where one finds $\beta^{\text{(MK)}}\approx 0.3047$, an error of roughly 31 \%. Additionally, we can compute critical exponents, which essentially confirm the results presented in \cite{kadanoff} on the Gu--Wen algorithm on the square lattice for the lowest cut--off. Exploiting the fact that decorated tensor networks preserve (some) of the original Ising spins, we can compute correlation functions. In figure \ref{fig:corr-len} we show the correlation functions (and its first derivative) of two `neighbouring' spins on a 2D lattice with periodic boundary conditions after different numbers of iterations, i.e. different distances between the spins on the original lattice. This correlation function clearly indicates the phase transition between the ordered and disordered phase, where the correlation functions intersect right on the phase transition because of scale invariance of the fixed point tensor.

Increasing the cut--off the phase transition temperature approaches the exact result; the highest cut--off tested is $\chi_{\mathrm{eff}} =60$, for which we observe a phase transition around $\beta \approx 0.440706$ (an error of roughly 0.06 \%).

For the 3D Ising model our method gives in the lowest order approximation a phase transition between $0.25 - 0.251$,
whereas Monte Carlo approaches \cite{montecarlo1,montecarlo2,montecarlo3} give around $\beta_c \approx 0.2217$.

\section{Discussion}

 Local gauge symmetry, and thus a redundancy of variables poses a serious challenge to tensor network algorithms aiming to efficiently coarse grain these systems and extract their effective dynamics at larger and larger scales. Yet their capability to also deal with negative and complex amplitudes, as encountered in fermionic systems and spin foams in quantum gravity, adds to their desirability.

Therefore we have introduced and tested a novel tensor network renormalization algorithm, called decorated tensor networks: Instead of encoding all variables into the tensor network we base our algorithms on building blocks of the discretisation, here cubes in the 3D cubical lattice, which are coloured by the original variables of the system and carry a tensor in their center; hence the tensor gets `decorated' by the original variables of the system.
This prevents a blurring of the gauge symmetry (e.g.\ represented in the form of Gauss constraints for the representation labels) which is otherwise bound to happen after several coarse graining steps and the variable redefinitions inherent to the tensor network algorithm. 

The decorated tensor network scheme allows for a clear identification and preservation of (part of) the original variables of the system. As an additional advantage the fixed points of this renormalization procedure are more straightforward to interpret. Note that the tensor network itself plays a distinctly different role in this algorithm: Instead of  encoding the full dynamics of the system, it rather appears as a way of locally encoding higher order corrections. To lowest cut--off, i.e. if all tensor indices are trivial, the systems keeps part of the original variables, either in untransformed form (in the case of the 2D Ising like models) or in a coarse grained form (in the case of Abelian lattice gauge theories, where the representation labels are coarse grained by using the Gau\ss~ constraints). %Higher order corrections can be understood as a higher multiplicity of particular colouring / configurations of the building blocks.

In fact, a slightly different perspective can be taken as well: the network of tensors, located in the centre of the building blocks, can be contracted at any point. After this contraction one arrives back at a system depending on the same type of variables as the original ones, yet with generically non--locally interacting building blocks. Hence one can understand the decorated tensor network as encoding non--local interactions between the building blocks, where the cut--off on the tensor indices can be directly translated into a truncation of the non--local interactions, see the discussions in \cite{dittcyl,review14}. % e.g. to lowest approximation, no non--local interactions will occur due to the trivial nature of the tensor network. 
This perspective will help to compare tensor network results with other, for instance continuum methods, based on truncations of (non--local) coupling terms in the effective action. 

In this vain we hope that decorated tensor networks might eventually facilitate new methods mixing both analytical (for the decoration indices) and numerical techniques (for the tensor indices). This can possibly provide new approaches, which might also allow to go beyond finite groups towards  Lie groups with infinite initial bond dimension and to allow to deal with (gauge) divergences occurring in spin foam model\cite{perini,aldo,bonzomdittrich}.

We tested the method with the Ising gauge model in 3D, which led to acceptable results. We are not aware of other results with tensor network algorithms applied to gauge theories. The method seems to be in particular effective for large structure groups (i.e. large initial bond dimension), which will appear in spin foams. One can define spin foam models with structure group given by a quantum group (at root of unity) \cite{qgroupmodels,qgroupmodels1,qgroupmodels2,qgroupmodels3,qgroupmodels4,qgroupmodels5,qgroupmodels6,qgroupmodels7}. This will lead to a finite initial bond dimension, which one would however like to push to larger values to emulate gravity with smaller cosmological constant.  Thus tensor network techniques are applicable to spin foams and provide in fact the first systematic coarse graining scheme (as Monte Carlo methods are not applicable). Lattice gauge theories with Lie groups as structure groups require a priori an infinite initial bond dimension. We discussed several strategies to deal with this in the introduction. Also here one can introduce quantum groups that would provide a (symmetry deforming) cut--off on the bond dimension. In this case one would like also to consider a family of models with larger and larger induced cut--off in order to reach the undeformed case.

We thus expect that one has to deal with a large initial bond dimension. Hence it is important to have an efficient method to achieve the lowest order approximation. We also presented and tested several possibilities to improve the lowest order approximations. We observed a systematic improvement of the phase transition temperature with increasing cut-off. Again it is difficult to compare with other tensor network methods -- general discussions suggest that implementation of entanglement filtering \cite{vidal,ferris} might improve very much the effectiveness of 3D algorithms.

A further remark, regarding the application to spin foams, is that the decorated tensor networks allow to keep the geometric interpretation of the spin foam variables. The coarse graining process might thus allow to conclude whether the blocking of variables (as encoded in so--called embedding or truncation maps \cite{dittcyl,review14}), determined by the SVD truncation, also proceeds in a geometric manner. As models of quantum gravity spin foams pose additional challenges. One is that the notion of scale, in terms of complexity of boundary data, as well as diffeomorphism symmetry have to be emergent \cite{dittrich08,dittrich12a,review14}. To this end non--local amplitudes are unavoidable \cite{measure1,measure2}, which indeed can be taken care off with the decorated tensor networks. As described in \cite{review14}  tensor networks  can be understood as a truncation scheme to compute physical states and a physical vacuum. We hope that the decorated tensor networks allow in particular to retain the geometric interpretation of the initial amplitudes, with the actual tensor network providing an improvement \cite{improved1,improved2} over the bare (initial) amplitudes.

\appendix

\section{Appendices}

\section{Non--Abelian lattice gauge theories}\label{app:lgt}

In this section we intend to briefly outline the differences in the representations -- and thus also tensor network representations -- for lattice gauge theories with non--Abelian structure groups in contrast to the Abelian case discussed in the main text, including a brief discussion on spin foam models (see the next section for an introduction, a brief discussion on coarse graining and references).

The partition function of non--Abelian lattice gauge theories is of the same form as \eqref{1}; group elements $g_e$ are assigned to the edges $e$ of the lattice and class functions $\omega_f$ sit on the faces $f$, with the property $\omega(h g h^{-1}) = \omega(g) \forall h,g$. The $\omega_f$ are evaluated on the face holonomy $h_f = \vec{\prod}_{e \subset f} g_e$, such that the partition function is invariant under local gauge transformations. These assign group elements $h_v$ to each vertex and transform the edge variables as $g_e \rightarrow h_{s(e)} g_e h_{t(e)}^{-1}$, where $s(e)$ and $t(e)$ denote source and target of the edge $e$ respectively.

As in the Abelian case, we can readily expand the class functions into characters $\chi_\rho$ of the irreducible representations $\rho$ of the structure group:
\ba \label{2a}
\omega(h)=\sum_\rho \tilde \omega(\rho) \chi_\rho(h) \, , 
\ea
however in contrast to the Abelian case, the character of a product of group elements splits into a product of representation matrices of group elements with contracted indices, $\chi_\rho(g_1 g_2) = \rho^m_{\;\;n}(g_1) \rho^n_{\;\;m}(g_2)$. After this transformation, we perform the group summation / integration and arrive at the following partition function:
 \ba \label{3a}
 Z &=& \sum_{\rho_f }  \prod_f \tilde \omega(\rho_f) \,   \prod_e ({\mathbf P}_e )^{\{n_f\}_{f \supset e}}_{ \{m_f\}_{f \supset e}} \left( \{\rho_f\}_{f \supset e} \right) \, .
 \ea

\begin{figure}
\begin{center}
\includegraphics[width=0.35\textwidth]{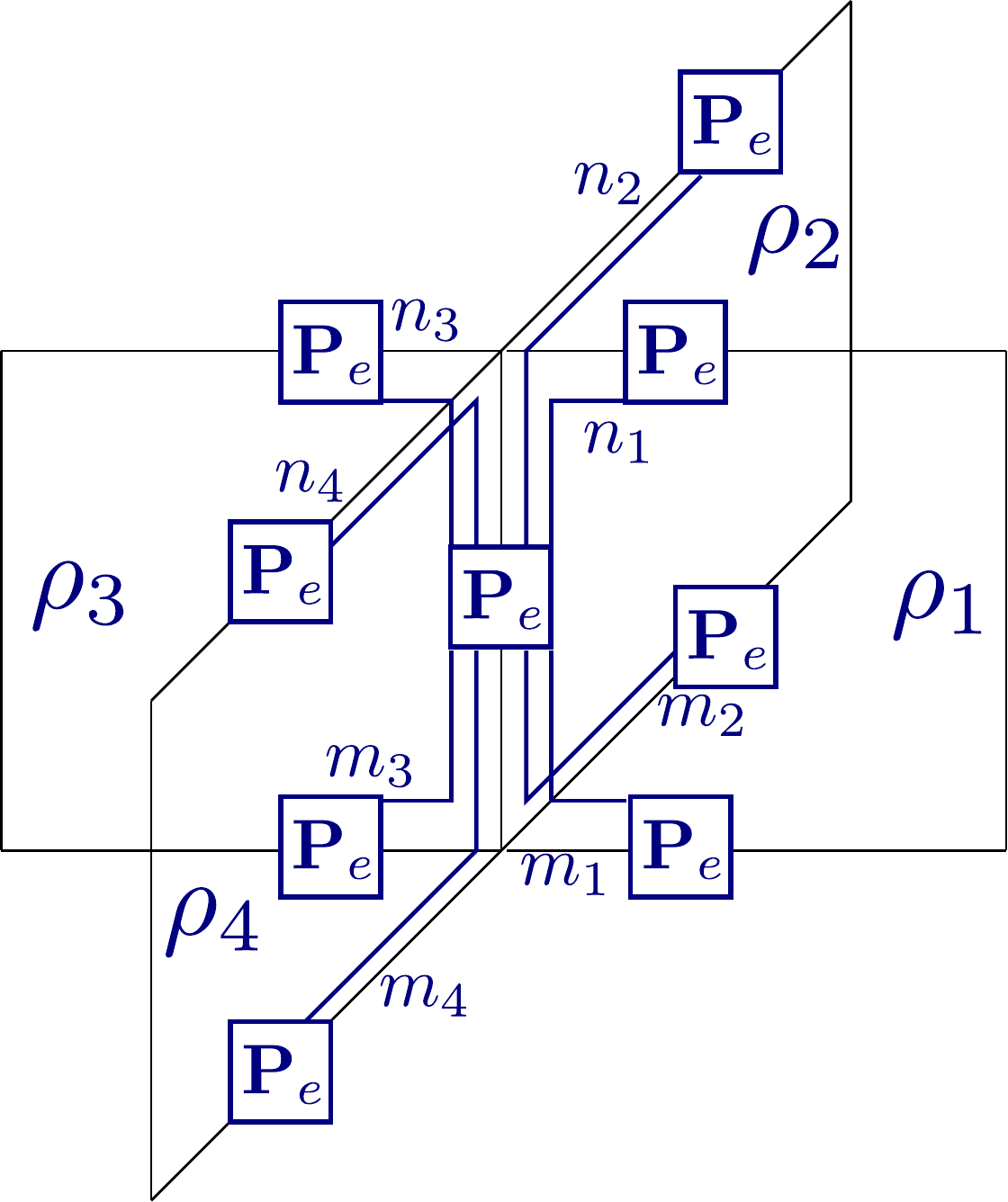}
\caption{Projector $\mathbf{P}_e$ is associated to each edge of the lattice. The magnetic indices are contracted with neighbouring projectors sharing the same face $f$. In case the orientations of the faces $f \supset e$ and the edge $e$ match, i.e. $o(e,f)=1, \, \forall f  \supset e$, then $\mathbf{P}_e : \text{Inv}\left(\bigotimes_{f \supset e} V_{\rho_f} \right) \rightarrow \text{Inv}\left(\bigotimes_{f \supset e} V_{\rho_f}\right) $,  the contragredient representation has to be used for opposing orientations. \label{fig:projector}}
\end{center}
\end{figure}

Note that this partition function differs from \eqref{AbZ} only in the replacements of the Gau\ss~constraints by the objects $\mathbf{P}_e$, the Haar projectors, that is representation theoretical objects associated to the edges of the lattice. These Haar projectors map onto the subspace associated to the trivial representation in the tensor product of representations $\rho_f$ associated to the faces adjacent to the edge $e$, that is the invariant subspace in the representation space $\otimes_{f \supset e} V_{\rho_f}$. The magnetic indices $m_e,n_e$ of $\mathbf{P}_e$ are contracted in a particular pattern, as shown in figure \ref{fig:projector}, arising from the splitting of characters into representation matrices.

The partition function for spin foam models actually is of the same form as \eqref{3a} with the slight difference that $\mathbf{P}_e$ is replaced by $\mathbf{P}'_e$, which is a projector onto a smaller subspace inside the trivial representation space. This is due to the fact that spin foam models implement additional conditions, so--called simplicity constraints, where a particular choice of these constraints leads to a particular choice of projector ${\mathbf P}'_e$. For further explanation see also \cite{finitesf,eckert1,eckert2}.

%Spin foam models implement simplicity constraints by replacing this tensor ${\mathbf P}_e$ by a projector ${\mathbf P}'_e$ onto a smaller subspace inside the trivial representation space, thus (\ref{3}) can serve also as the partition function for spin foam models (for further explanation see for instance \cite{finitesf,eckert1,eckert2}). The particular choice of simplicity constraints is then encoded into a particular choice of projector ${\mathbf P}'_e$.

%For Abelian structure groups, for instance ${\mathbb Z}_K$, the partition function (\ref{3}) simplifies to (we will denote the representations for ${\mathbb Z}_K$ by $k \in 0,\ldots, K-1$)
%\ba
%Z&=&\sum_{k_f}  \prod_f \tilde \omega(k_f)  \,   \prod_e  \delta\left(\sum_{f \supset e} o(e,f) \, k_f\right) \, ,
%\ea
%where $o(e,f)=-1$ if face and edge orientation disagree and $o(e,f)=+1$ if these agree. Thus the Haar projectors reduce to Gauss constraints, imposing that the representations around the faces meeting in one edge sum up to zero. 

%\begin{figure}
%\begin{center}
%\includegraphics[width=0.39\textwidth]{figures/sf-tensor}
%\caption{Tensor network representation of a spin foam defined on a cubical lattice. The tensors $T_e$ on the edges capture the intertwiner degrees of freedom, while the tensors $T_f$ on the faces are an auxiliary structure. For a given face $f$ they ensure that all edge tensors $T_e$ with $e \subset f$ carry the same representation label $\rho_f$ assigned to that face.
%\label{fig:sf-tensor}}
%\end{center}
%\end{figure}

Analogous to the Abelian case, this representation can be cast into a tensor network with the tensors $\mathbf{P}_e$ on the edges and auxiliary tensors on the faces ensuring the same representation $\rho_f$ is seen by all tensors bounding that face. However, this tensor network representation not only suffers from saving redundant data as in the Abelian case but also from the complicated contraction scheme of the magnetic indices prescribed by the gauge invariance of the model. Instead, one would prefer a representation in which these indices are pre--contracted as in the symmetry protecting algorithms developed in \cite{merce,qgroup} (for models with global symmetry).
%Although we have now variables associated to faces, this form of the partition function (\ref{3}) can be brought into the form of a tensor network. To this end one associates tensors $({\mathbf P}_e )$ to the midpoints of each edge and introduces auxiliary tensors to the midpoints of the faces, that make sure, that each tensor $({\mathbf P}_e )$ around a given face $f$, sees the same representation label $\rho_f$, see also figure \ref{fig:sf-tensor}. This representation appeared in \cite{eckert1,eckert2} and also later in  \cite{meurice}.
%
%However, this representation (\ref{3}) is rather inefficient if it comes to coarse graining algorithms. The dependence of the tensor  in the magnetic indices is prescribed by the gauge invariance of the model, and thus one would rather wish to pre--contract these indices, as in fact happens for the symmetry protecting algorithms developed in \cite{merce2,qgroup} (for models with global symmetry). 
Thus one would only have to deal with representation and possibly intertwiner labels (which for multiplicity free\footnote{I.e. groups in which the tensor product of two irreducible unitary representations reduces to a sum in which every irreducible unitary representation appears at most once.} groups are again representation labels), as in \cite{merce,qgroup}. This would also allow to understand the coarse graining in terms of weights on intertwiners.

\begin{figure}
\begin{center}
\includegraphics[width = 0.39 \textwidth]{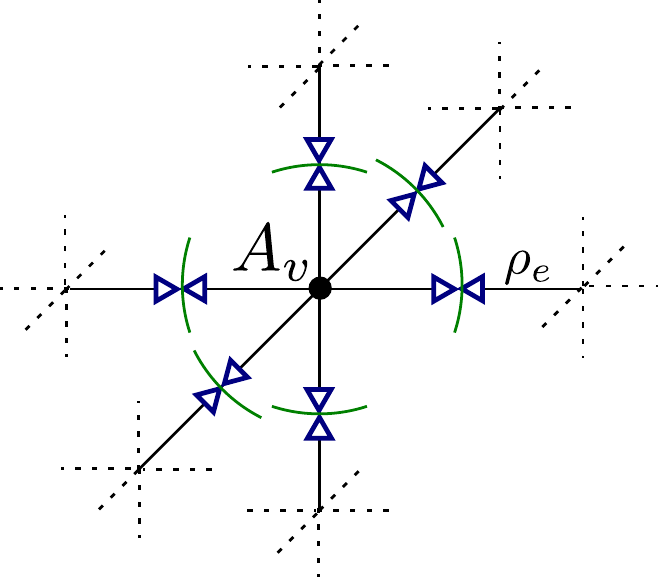}
\caption{Construction of the vertex amplitude: The projectors $\mathbf{P}_e$ on the edges are expanded in the intertwiner basis $\{\iota_e\}$ according to \eqref{4}, the opposing triangles pictorially representing $\left | \iota_e \right \rangle \left \langle \iota_e \right |$. Together with the contraction of magnetic indices shown in figure \ref{fig:projector}, the projectors can be split and associated to the vertex of the lattice.
\label{fig:vertex-amp}}
\end{center}
\end{figure}

Fortunately, this goal can be achieved by rewriting the partition function in terms of vertex amplitudes as in the Abelian case, yet with the complication that the intertwiners of non--Abelian groups are non--trivial. A standard technique in spin foams is to split the projector ${\mathbf P}_e $ -- by using an orthonormal basis $|\iota_e\rangle$ of the invariant subspace of $\otimes_{f \supset e} V_{\rho_f}$  -- into two parts (referred to as intertwiners)\footnote{We assume here for simplicity that all faces adjacent to the edge $e$ have an orientation that agrees with the one of $e$.}
\ba\label{4}
 ({\mathbf P}_e )^{\{n_f\}_{f \supset e}}_{ \{m_f\}_{f \supset e}}  = \sum_{\iota_e}  {}^{\{n_f\}_{f \supset e}}\left| \iota_{e} \right \rangle \left\langle \iota_{e} \right |_{ \{m_f\}_{f \supset e}} \, ,
\ea
see figure \ref{fig:vertex-amp}.
These two parts can be associated to the source vertex and target vertex of the edge $e$ respectively. The sum is over intertwiner labels $\iota_e$ (which are again representation labels assuming multiplicity free groups) appearing in the tensor product $\otimes_{f \supset e} V_{\rho_f}$. 

Now every half edge, labelled by an edge--vertex pair $(e,v)$, carries an object $|\iota_e\rangle$. The structure of the magnetic index contraction is such, that the sum over the magnetic indices factorizes over the vertices, i.e. the set of objects $\iota_{e}$ associated to a given vertex $v$ can be contracted to the vertex amplitude $A_v\left( (\rho_f)_{f \supset v}, (\iota_e)_{e \supset v}\right)$, similar to the one in the Abelian case in \eqref{AbAv}, again with the difference that it also depends on intertwiner labels $\iota_e$. The partition function is then given by:
%  that only depends on representation labels associated to the faces and edges adjacent to the vertex $v$. This is shown in figure \ref{fig:A_v}. Thus we have now a form of the partition function
\ba\label{5}
Z &=& \sum_{\rho_f, \iota_e} \prod_f \tilde \omega_f \prod_v A_v\left( (\rho_f)_{f \supset v}, (\iota_e)_{e \supset v}\right) \, .
\ea
%which almost is a tensor network, if it would not be for the representation labels on the faces. (The face weights can be absorbed into the vertex weights for a regular lattice.)

\begin{figure}
\begin{center}
\includegraphics[width = 0.35 \textwidth]{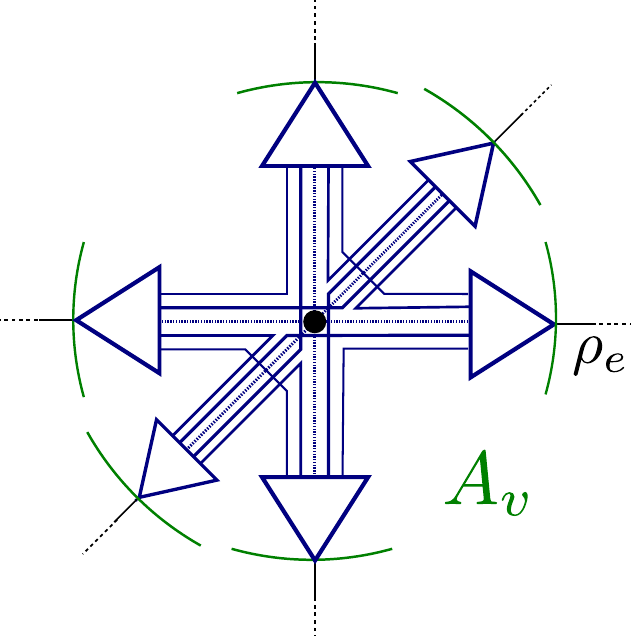}
\caption{The construction of the vertex amplitude $A_v$ for a 3D cubical lattice with the explicit contraction of magnetic indices.
\label{fig:A_v}}
\end{center}
\end{figure}

However as in the Abelian case, writing this system as a tensor network either requires again the introduction of auxiliary tensors on the faces or the copying of representations on the faces onto the edges with additional Kronecker deltas in the vertex plus an additional intertwiner label $\iota_e$ per edge. This would mean that we initially assign $6 \times 5$ representation labels to this edge for a cubical 3D lattice, with the restriction that the four representation labels $\rho_f$ actually couple to the intertwiner $\iota_e$. 

Thus also in this case it is more efficient to encode a non--Abelian lattice gauge theory or spin foam model into a decorated tensor network. We sketched a coarse graining algorithm based on such a decorated tensor network in the main text.

\section{Spin foams} \label{app:spinfoams}

One of the main motivations for developing the algorithm presented here is the aim to coarse grain spin foam models. In the following we will briefly outline the relevance of tensor network renormalization in realizing this aim.

Spin foams \cite{foams1,foams2,foams3} are generalized versions of lattice gauge theories, but feature a more complicated structure. Roughly speaking, the models are parametrized by weights for intertwiners.  This means that the weights  depend on a combination of several representation labels, instead of being parametrized by face weights, which are functions of one representation label only and describe standard lattice gauge models. %the main difference to non--Abelian lattice gauge theories is discussed in appendix \ref{app:lgt}. 
 Also due to these more complicated amplitudes the phase structure of spin foams, beyond those phases that also appear in lattice gauge theory, is basically unknown, except for the recent developments in \cite{merce,holonomy1,qgroup}. Tensor network algorithms are indeed so far the only tool that is available for the real space coarse graining of these systems (as even the Migdal Kadanoff relations \cite{MK1,MK2} do not apply to spin foams and Monte Carlo simulations are not applicable due to complex amplitudes). Thus a development of these techniques is worthwhile, although there are still some general issues about tensor network techniques to address  \cite{cdl,ferris,kadanoff}.

The full spin foam models are defined in four dimensions and involve a structure group $SU(2)\times SU(2)$ (this is for models describing Euclidean signature metrics, which face less technical challenges than those describing Lorentzian metrics, based on $SO(3,1)$). However, these models are barely understood beyond a very few building blocks, additionally the issue of divergences arises \cite{perini,aldo,bonzomdittrich}. \cite{finitesf,eckert1,eckert2} therefore proposed a program in which one considers a reduction of the structure group to a finite group, and a dimensional reduction to two dimensional spin net models. Nevertheless a key input of spin foam construction, the simplicity constraints \cite{foams1,foams2,foams3,holonomy1}, and with it, the emphasis on the behaviour of intertwiner degrees of freedom \cite{merce,qgroup}, was kept in these models. In \cite{qgroup} spin net models based on a quantum group $SU(2)_k$ are considered which, at least in 3D, describe gravity with a cosmological constant. In this sense  the work  \cite{qgroup} lifted the reduction of the structure group and revealed a very rich phase space structure for quantum group spin nets, related to phases in anyonic spin chains \cite{anyonic1,anyonic2,wojtek}. Spin net models can be in fact interpreted as spin foams with a large number of (dual) edges and two vertices \cite{qgroup}. The different phases describe the effective coupling between the two spin foam vertices, where this coupling is determined by the (initial) choice of simplicity constraints.

It is now crucial to investigate the phases for spin foams in order to test the conjecture that spin nets could indeed possess a phase structure similar to spin foams (this is similar to the relation between lattice gauge theories and Ising like models defined for the same group \cite{kogut}). To this end it is useful to incorporate techniques that have been developed for spin nets \cite{merce,qgroup} in (a) dealing with (global or gauge) symmetries and (b)  allowing for a monitoring of the weights for the intertwiner degrees of freedom during the coarse graining. This was accomplished by a symmetry protecting algorithm, in which magnetic indices are pre--contracted, see also \cite{vidalsymm} and the tensor decomposes in blocks labelled by intertwiner channels. The decorated tensor network algorithm presented here allows indeed for similar mechanisms.

\section{Acknowledgements}

The authors thank Leo Kadanoff for an enlightening discussion. S.M. would like to thank Perimeter Institute for hospitality during this work and Girton College, Cambridge for support. S.St. would like to thank Perimeter Institute for an Isaac Newton Chair Graduate Research Scholarship. This research was supported by Perimeter Institute for Theoretical Physics. Research at Perimeter Institute is supported by the Government of Canada through Industry Canada and by the Province of Ontario through the Ministry of Research and Innovation. 

\bibliography{biblio}

\end{document}